\documentclass[aps,prl,twocolumn,showpacs,amsmath,floatfix]{revtex4}

\usepackage[utf8]{inputenc} 
\usepackage[T1]{fontenc}    
\usepackage{hyperref}       
\usepackage{url}            
\usepackage{booktabs}       
\usepackage{amsfonts}       
\usepackage{nicefrac}       
\usepackage{microtype}      

\usepackage[]{amsmath}
\usepackage{amssymb}
\usepackage[dvips]{graphicx}
\usepackage{color}
\usepackage{tabularx}
\usepackage{mathtools}
\usepackage{algpseudocode}
\usepackage{enumitem}
\usepackage{multirow}
\usepackage{epstopdf}  
\usepackage{bm}
\usepackage{floatrow}

\usepackage{footnote}
\makesavenoteenv{tabular}

\newcommand{\vect}[1]{\textbf{\textit{#1}}}

\begin{document}

\title{
Deep neural network for the dielectric response of insulators
}
\author{Linfeng Zhang}
\affiliation{Program in Applied and Computational Mathematics, Princeton University, Princeton, NJ 08544, USA}
\author{Mohan Chen}
\affiliation{CAPT, HEDPS, College of Engineering, Peking University, Beijing 100871, China}
\author{Xifan Wu}
\affiliation{Department of Physics, Temple University, Philadelphia, PA 19122, USA}
\author{Han Wang}
\email{wang_han@iapcm.ac.cn}
\affiliation{Laboratory of Computational Physics, Institute of Applied Physics and Computational Mathematics, Huayuan Road 6, Beijing 100088, China}
\author{Weinan E}
\affiliation{Department of Mathematics and Program in Applied and Computational Mathematics, Princeton University, Princeton, NJ 08544, USA}
\affiliation{Beijing Institute of Big Data Research, Beijing, 100871, P.R.~China}
\author{Roberto Car}
\email{rcar@princeton.edu}
\affiliation{Department of Chemistry,  
Department of Physics, 
Program in Applied and Computational Mathematics, 
Princeton Institute for the Science and Technology of Materials,  
Princeton University, Princeton, NJ 08544, USA}

\begin{abstract}
We introduce a deep neural network to model in a symmetry preserving way the environmental dependence of the centers of the electronic charge. 
The model learns from {\it ab initio} density functional theory, wherein the electronic centers are uniquely assigned by the maximally localized Wannier functions. 
When combined with the Deep Potential model of the atomic potential energy surface, the scheme predicts the dielectric response of insulators for trajectories inaccessible to direct {\it ab initio} simulation. 
The scheme is non-perturbative and can capture the response of a mutating chemical environment.  
We demonstrate the approach by calculating the infrared spectra of liquid water at standard conditions, and of ice under extreme pressure, when it transforms from a molecular to an ionic crystal.     
\end{abstract}

\maketitle
Machine learning (ML) schemes introduced in the last decade~\cite{behler2007generalized, bartok2010gaussian,rupp2012fast,montavon2013machine,botu2016machine,chmiela2017machine,schutt2017schnet,smith2017ani,han2017deep,zhang2018deep,zhang2018end}   can model accurately the potential energy surface (PES) of a multi-atomic system upon training with first-principle electronic density functional theory (DFT) data~\cite{kohn1965self}.
These approaches extend the size and time range of {\it ab initio} molecular dynamics (AIMD)~\cite{car1985unified,marx2009ab}, and make possible studies of rare events, such as crystal nucleation, with enhanced sampling methodologies in simulations of {\it ab initio} quality~\cite{bonati2018silicon}.

Most approaches, so far, focused on representing the dependence of the PES, a scalar quantity, on the atomic coordinates.
Recently, methods to fit the environmental dependence of electronic properties have been proposed~\cite{gastegger2017machine,grisafi2018symmetry,grisafi2018transferable,wilkins2019accurate,chandrasekaran2019solving,zepeda2019deep,raimbault2019using,kapil2019inexpensive}.
In particular, kernel based methods have been used to represent the polarization and its time derivatives~\cite{grisafi2018symmetry,raimbault2019using,kapil2019inexpensive}, which are needed in many studies of materials,
including calculations of infrared (IR)~\cite{sharma2005intermolecular}, Raman~\cite{putrino2000generalized,wan2013raman}, and sum frequency generation (SFG)~\cite{wan2015first} spectra, transport calculations in ionic liquids~\cite{rozsa2018ab} and superionic crystals~\cite{sun2015phase,wood2006dynamical,schwegler2008melting}, and simulations of ferroelectric phase transformations~\cite{srinivasan2003pbtio3,fluri2017enhanced}.
These schemes learn from AIMD trajectories, but, so far, the calculated spectra of liquid water~\cite{kapil2019inexpensive} do not match the quality of direct many-body expansions of the dipole moment~\cite{liu2015quantum}.

Here we propose an alternative approach based on deep neural networks (DNNs) and maximally localized Wannier functions (MLWFs)~\cite{marzari1997maximally,marzari2012maximally}, i.e. electronic orbitals with minimal spatial spread obtained from a unitary transformation of the occupied orbitals in insulators. 
In spin saturated systems the MLWFs describe electron pairs. 
Thus, upon assigning charges of 2$e^-$ to the Wannier centers (WCs), i.e. the centers of the MLWF distributions, the electric polarization is the dipole moment 
$\bm{M}$ of the neutral system of point charges made by the WCs and the atomic nuclei.
In extended periodic systems, $\bm{M}$ includes an arbitrary quantum~\cite{Resta1994RMP}, but its derivatives and correlation functions are well defined and describe observable properties.

In systems with different atomic species, the WCs  fluctuate near the most electronegative atoms during molecular evolution. 
As a consequence of the nearsightedness of the electronic matter~\cite{kohn1996density, prodan2005nearsightedness}, the WCs only depend on the atoms in their local environment, 
and their positions can be accurately represented by the DNN model, called Deep Wannier (DW), which is introduced here.
It is an end-to-end scheme that does not use any {\it ad hoc} construction, in addition to the coordinate information and the network itself, to map input atomic coordinates into output WC coordinates. 
The model is size extensive, preserves translational, rotational, and permutational symmetry, and yields a polarization that varies continuously with the atomic coordinates. 
As such it can describe how the polarization responds to chemical bond changes in electronic insulators.

\begin{figure}
  \centering
  \includegraphics[width=0.60\textwidth]{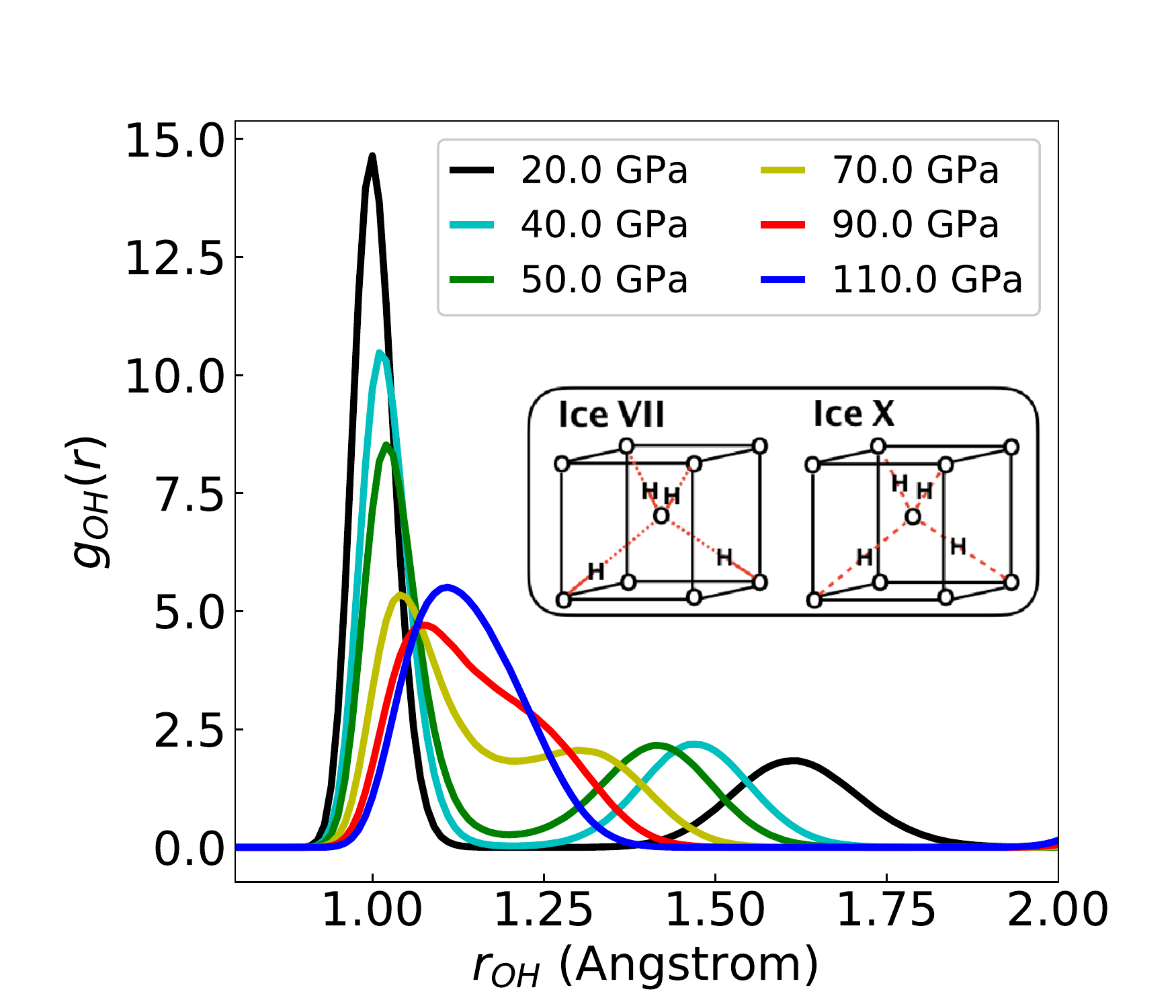}
  \caption{(Color online) Schematic illustration of the transitions from ice VII to ice VII' and from ice VII' to ice X by the O-H pair correlation functions $g_{OH}(r)$ calculated at $T$=300 K and various pressures. Inset: Idealized structure of ice VII and ice X.
      \label{fig:rdf-ice}}
\end{figure}

To demonstrate the methodology, we compute (a) the IR absorption spectrum of liquid water at standard temperature and pressure, and (b) the frequency dependent imaginary dielectric function in the high pressure transformation of ice VII to ice VII', and from ice VII' to ice X~\cite{hernandez2018proton}.
In ice VII there are 2 donor and 2 acceptor H-bonds per oxygen, as stipulated by the ice rule~\cite{bernal1933theory,pauling1935structure}, and the water molecules are always well defined. 
In ice VII' the hydrogens hop between two equivalent sites near each oxygen of an O-O bond, occasionally violating the ice rule.
In ice X the hydrogens sit in the mid O-O bonds, 
the water molecules cannot be identified,
and the system is better described in terms of O${}^{2-}$ and H${}^+$ ions. 
This behavior is reflected in the O-H pair correlation functions in Fig.~\ref{fig:rdf-ice}. 
For more sophisticated analyses and a phase diagram see e.g. Ref~\cite{hernandez2018proton}.      
The polarization change accompanying the above transformations is seamlessly described by DW. 
The scheme should also work at higher temperature when ice X becomes superionic~\cite{millot2019nanosecond}, or at even higher temperature when the superionic crystal melts into an ionic liquid~\cite{rozsa2018ab}. 
Similarly, the scheme should work for proton transfer events in water at standard conditions when neutral molecules interconvert with hydronium and hydroxide ion complexes (see e.g. Ref.~\cite{chen2018hydroxide}).  

In our implementation, DW is trained with valence only pseudopotential electronic structure calculations.
In the laboratory frame, the positions of the atoms and of the WCs are  $\bm{r}_1, \bm{r}_2 , ... ,\bm{r}_i , ...,\bm{r}_N$ and $\bm{w}_1, \bm{w}_2 , ..., \bm{w}_i , ...,\bm{w}_{N_w}$, respectively. 
We assume, for simplicity, that the WCs are only associated to one atomic species and consider water, in which there are 4 WCs per oxygen, as a concrete example
\footnote{Generalizations to more than one reference atom would be feasible to deal with situations in which the WC are shared among covalently bonded atoms of same electronegativity.}. 
It is easy to select with a cutoff distance the 4 WCs with coordinates  $\bm{w}_{l_i =1,..,4}$ that are closer to the oxygen $i$ located at $\bm{r}_i$. 
Their centroid
\begin{equation}
 \bm{w}_{i} =\frac{1}{4}\sum_{l_i =1}^4 \bm{w}_{l_i}
  \label{eqn:w-i}
\end{equation}
is well defined even when water molecules cannot be identified. 
Our aim is to construct a vector function $\bm{f}$ that gives $\bm{w}_{i}$ if the positions of the $N_i$ atoms $\bm{r}_k$
in the neighborhood $\mathcal{N}_i$ of $\bm{r}_i$, defined by the radius $r_c$, are known, i.e.:
\begin{equation}
 \bm{w}_{i}=\bm{f}(\{\bm{r}_{k},k\in\mathcal{N}_i\}).
  \label{eqn:w-i-NN}
\end{equation}
Here $i$ is an oxygen atom but the atoms $k$ include oxygens and hydrogens. 
To ensure that $\bm{f}$ preserves the translational, rotational, and permutational symmetry, we generalize the scheme for the Deep Potential (DP), the DNN representing the PES, introduced in Refs~\cite{zhang2018deep,zhang2018end}.

First, we make a $local$ frame transformation to the primed coordinates, which preserve translational symmetry:
\begin{equation}
\bm{r}_{ki}'\equiv\bm{r}_k-\bm{r}_i, \quad\bm{w}_i'\equiv\bm{w}_i-\bm{r}_i.
  \label{eqn:r-w-local}
\end{equation}
Then, we introduce a weight function $s(r_{ki}')$ equal to $1/r_{ki}'$ at short distance $r'_{ki}=(\bm{r}_{ki}'\cdot\bm{r}_{ki}')^{1/2}$, and decaying smoothly to zero as $r_{ki}'$ approaches $r_c$. 
Using $s(r_{ki}')$, we describe the atomic coordinates $\bm{r}_{ki}'=(x_{ki}', y_{ki}', z_{ki}')$ with the 4-vector $\bm{q}_{ki} =(q_{ki}^1, q_{ki}^2, q_{ki}^3,q_{ki}^4)$$=(s(r_{ki}'), s(r_{ki}')x_{ki}', s(r_{ki}')y_{ki}', s(r_{ki}')z_{ki}')$ to enforce continuous evolution when atoms enter/exit the neighborhood.  

Next, we enforce permutational invariance and rotational covariance by introducing two DNNs, an $embedding$ DNN and a $fitting$ DNN.
The number of hidden layers and outputs is refined in the training procedure.
The embedding DNN is the matrix $\bm{E} = (E_{ik}) = ( E_i (s(r_{ki}') ))$ with $M$ rows and $N_i$ columns optimized by training, which maps the set $\{s(r_{ki}'), ~k \in \mathcal{N}_i \}$ onto $M$ outputs.

The set of generalized coordinates $\{\bm{q}_{ki}\}$ in a neighborhood is represented by the matrix $\bm{Q}= (Q_{k\lambda}) = (q_{ki}^\lambda)$ with $N_i$ rows and 4 columns.
Multiplication of $\bm{E}$ by $\bm{Q}$ gives the matrix $\bm{T}=\bm{E}\bm{Q}$ with $M$ rows and 4 columns, whose generic element is:  
\begin{equation}
T_{i\lambda}=\sum_{k=1}^{N_i}E_i(s(r_{ki}'))Q_{k\lambda}.
  \label{eqn:Tij}
\end{equation}
Let $\bm{S}$ be the matrix formed by the first $M'$ (<$M$) rows of $\bm{T}$. 
Multiplication of $\bm{T}$ by $\bm{S}^T$, the transpose of $\bm{S}$, gives the matrix $\bm{D}$ of dimension $M\times{M'}$, called the {\it{feature}} matrix:
\begin{equation}
\bm{D}=\bm{T}\bm{S}^T.
  \label{eqn:D}
\end{equation}
$\bm{D}$ is the argument of the fitting DNN, a row matrix $\bm{F}(\bm{D})=\{F_j(\bm{D}),j=1,...,M\}$ that converts the atomic coordinate information encoded in $\bm{D}$ onto $M$ outputs, 
which are mapped onto the centroid $\bm{w}_i'=(w'{}_i^1,w'{}_i^2,w'{}_i^3)$ upon multiplication with with the last three columns of $\bm{T}$:
\begin{equation}\label{eqn:wp}
w'{}_i^\lambda=\sum_{j=1}^{M}F_j(\bm{D})T_{j,\lambda+1}.
\end{equation}
Finally, $\bm{w}_i$ is retrieved from $\bm{w}_i'$ using Eq.~\ref{eqn:r-w-local} and one obtains the desired representation of Eq.~\ref{eqn:w-i-NN}.

We notice that $\bm{w}_i$ constructed in the way introduced above naturally preserves all the symmetry requirements.
Translational symmetry is preserved by the adoption of a local frame and relative positions in Eq.~\ref{eqn:r-w-local}.
Permutational symmetry is preserved by the smooth sum over the neighboring atoms in Eq.~\ref{eqn:Tij}.
Finally, as shown in Eq.~\ref{eqn:Tij}, the last three columns of $\bm{T}$ ($\lambda=2,3,4$) transform covariantly under rotation because $(Q_{k2}, Q_{k3}, Q_{k4})$ transforms like $\bm{r}_{ki}'$.
Then it is straightforward to verify that the elements of $\bm{D}$, and hence $\bm{F}(\bm{D})$, are invariant under rotation.
Therefore,  $w'{}_i^\lambda$ in Eq.~\ref{eqn:wp} transforms like $T_{j,\lambda+1}$, and is hence rotationally covariant.
 The values of $M$, $M'$, and of the number of layers of the DNNs are chosen empirically based on performance. 
In the applications discussed in this paper we adopt $M$=100 (of the same order of the number of atoms in a neighborhood), $M'$=6, and use a 3-layer representation for all the DNNs.  
The parameters $\bm\gamma$ of the embedding and fitting networks are determined by training, i.e., an optimization process that minimizes a loss function, 
here the mean square difference between the DW prediction and the training data. 
The Adam stochastic gradient descent method~\cite{Kingma2015adam} is adopted for the optimization.
Generalization of this formalism to tensor properties (like the polarizability) is introduced in Ref.~\cite{grace2020raman}.
 
DW should be combined with a DNN for the PES to study the evolution of the polarization along MD trajectories. 
For consistency, the two networks should be trained with electronic structure data at the same level of theory,
as in the applications below, which used DW and the DP representation of the PES~\cite{zhang2018deep,zhang2018end}. 
{\it Ab initio} electronic structure data are expensive and efficient learning strategies are crucial. 
We used the iterative learning scheme of Ref.~\cite{zhang2019active}. 
In this approach, a DNN, initially trained with a limited pool of {\it ab initio} data, is used to explore inexpensively the configuration space. 
A small subset of the visited configurations is selected with a suitable error indicator and single shot {\it ab initio} calculations are performed at these configurations. 
Training with the new data improves the model for further exploration and selection, followed by new data acquisition and learning. 
The protocol is repeated until all the explored configurations are described with satisfactory accuracy. 
The error indicator exploits the highly non-linear dependence of the DNN models on the network parameters. 
As a consequence, different initializations of the parameters lead to different local minima in the landscape of the loss function, originating an ensemble of minimizing DNNs. 
The variance of the predictions within this ensemble is an intrinsic property of a DNN model and is often a reliable indicator of its accuracy~\cite{podryabinkin2017active,zhang2018reinforced}. 
In our experience, good DNN models constructed with the above procedure require significantly less {\it ab initio} data in the target thermodynamic range than learning approaches based on independent AIMD sampling data.   

In the following we report calculations on liquid water at STP and on ice undergoing pressure induced structural phase transitions. 
For liquid water at STP, we did not use the incremental data generation scheme, since the training data were available and accessible online from previous work~\cite{ko2019isotope}.
For high-pressure ice, electronic structure data were not available, and we constructed DP and DW from scratch using the incremental learning procedure outlined above.
Full details on the implementation, training, and validation of the models (DP and DW) are given in the Supplemental Material [SM].
The code for this work has been integrated into the open-source software package DeePMD-kit~\cite{wang2018kit} and we used the DP-GEN package~\cite{zhang2019dpgen} for the iterative scheme.

\begin{figure}
  \centering
  \includegraphics[width=1.0\textwidth]{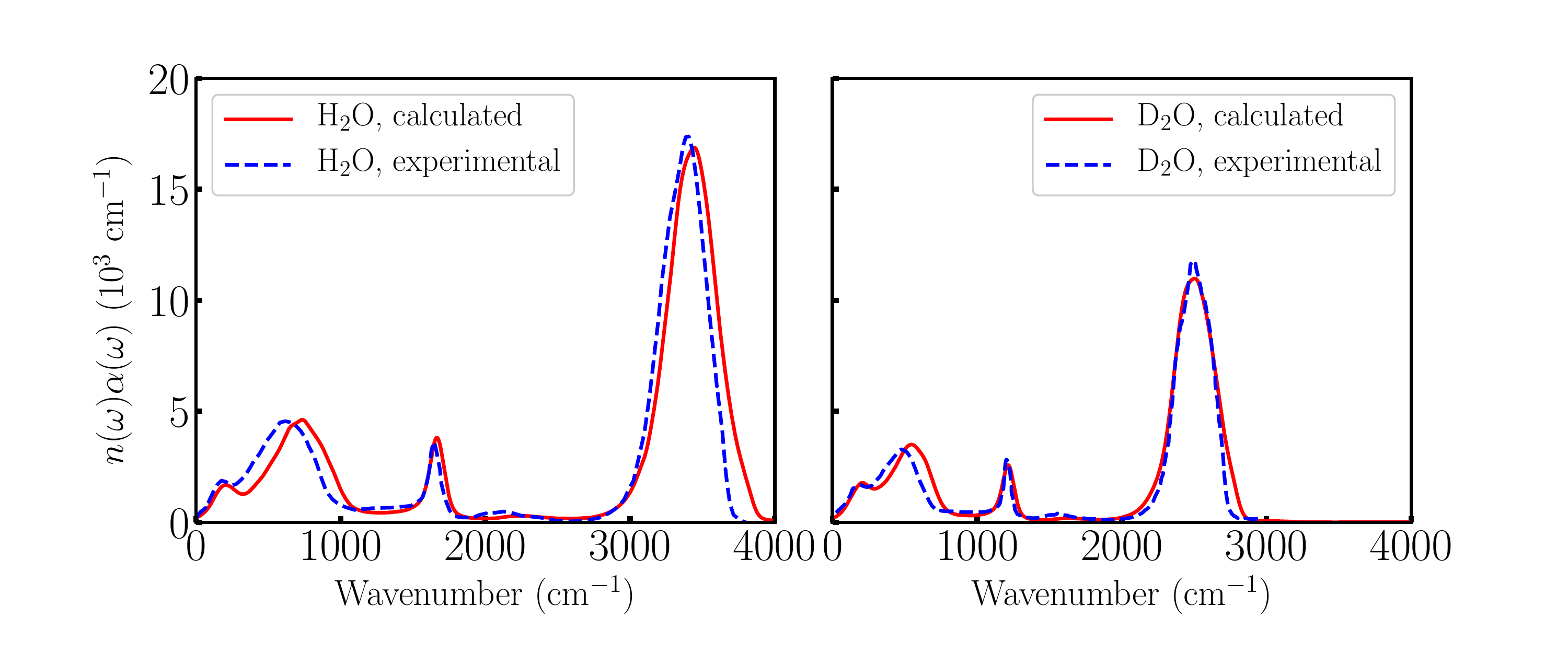}
  \caption{(Color online) IR absorption spectra of liquid H${}_2$O (left) and D${}_2$O (right). 
  The continuous lines report DPMD+DW calculations at $\sim$300 K. 
  The dashed lines report experimental data for H${}_2$O at 298 K (Ref.~\cite{bertie1996infrared}) and for D${}_2$O at 295 K (Ref.~\cite{bertie1989infrared}).
  \label{fig:ir}}
\end{figure}

We use DFT at the hybrid functional level (PBE0~\cite{Carlo1999PBE0}) with dispersion corrections~\cite{TS2009TS} for STP water.  
Using DW and DP~\cite{zhang2018end} we calculate the IR absorption spectra of liquid H${}_2$O and D${}_2$O for a cell with 512 molecules under periodic boundary conditions. 
We use two microcanonical trajectories lasting 0.5 ns each, for H${}_2$O and D${}_2$O, at an average temperature of $\sim$300 K at the equilibrium density of the simulation.
The frequency dependent absorption coefficient per unit length, $\alpha(\omega)$, times the refractive index, $n(\omega)$, is given by the Fourier transform of the time correlation function of the time derivative of the cell polarization $\dot{\vect{M}}=+6e\sum_l\dot{\bm{r}}_{O_l}+e\sum_m\dot{\bm{r}}_{H_m}-8e\sum_{n}\dot{\bm{w}}_{n}$ according to: 
\begin{align}\label{eqn:ir}
\alpha(\omega)n(\omega)=\frac{2\pi\beta}{3cV}\int_{-\infty}^{+\infty}dte^{-i\omega{t}}\langle\dot{\vect{M}}(0)\cdot\dot{\vect{M}}(t)\rangle,
\end{align}
where $V$ is the volume, $\beta = 1/k_BT$ is the inverse temperature, and $k_B$ is Boltzmann's constant.
Fig. ~\ref{fig:ir}  shows that the calculated spectra are in good agreement with the corresponding experimental observations. 

Similarly accurate IR spectra of liquid water can be obtained from representations of the PES~\cite{babin2013development,shank2009accurate} and the dipole moment~\cite{liu2015quantum} based on many-body molecular expansions. 
These powerful approaches are limited to molecular liquids and crystals. 
By contrast, our non-perturbative method works also for non-molecular systems, as we demonstrate by considering ice at T=300 K in the pressure range from 20 to 110 GPa, wherein structural phase transitions from ice VII to ice VII' and to ice X occur. 
We adopted the PBE functional approximation of DFT as in Refs.~\cite{schwegler2008melting,hernandez2018proton}.
We constructed DP/DW networks for ice in the temperature range from 240 to 330 K and pressure range from 20 to 120 GPa with the iterative learning approach set forth earlier in this paper. 
This procedure required a total of 2248 single shot DFT calculations with a 16 molecule cell and a total of 2400 single shot DFT calculations with a 128 molecule cell. 
For each cell size, the corresponding computational effort was less than the cost of a short ($\sim$5 ps) AIMD trajectory.  

We sampled the ice configurations at 300 K in the pressure range 20 - 110 GPa with constant pressure DPMD on a variable periodic cell with 128 water molecules, using a mild Nos{\'e}-Hoover thermostat~\cite{nose1984unified,hoover1985canonical} with damping time of 5 ps, much longer than the vibrational periods, to control the temperature.
Using DP+DW, we calculated $\alpha(\omega) n(\omega)$ according to Eq.~\ref{eqn:ir} with a set of 0.5 ns long trajectories at various pressures.
A direct AIMD study of the spectral changes in the closely related transformations from ice VIII (the proton ordered form of ice VII) to ice VII' and to ice X, reported in a pioneering paper by Bernasconi et al.~\cite{bernasconi1998ab}, used a small 16 molecule cell and $\sim$10 ps long trajectories. 

\begin{figure}
  \centering
  \includegraphics[width=0.80\textwidth]{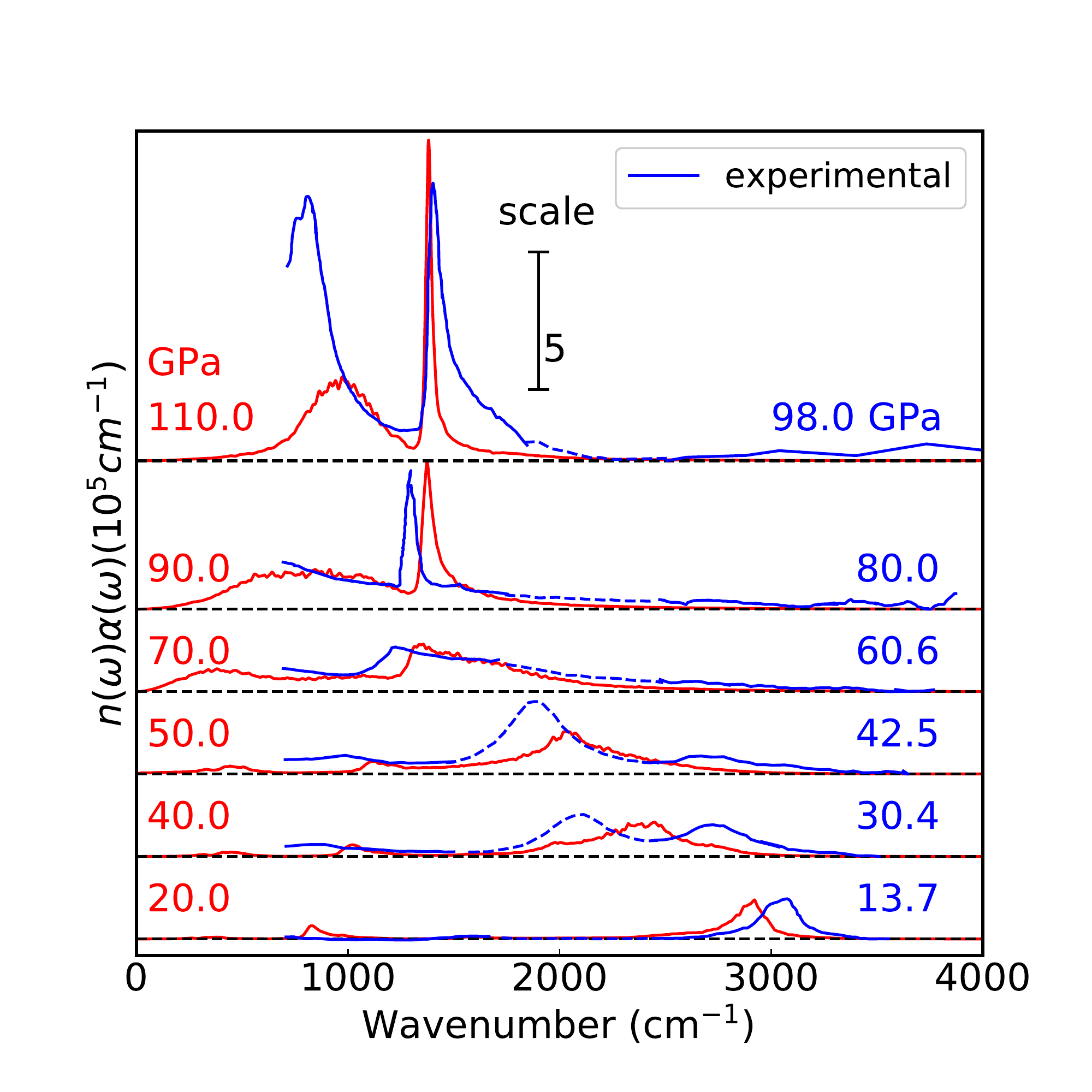}
  \caption{(Color online) Red: IR absorption spectra of H${}_2$O ice at 300 K and pressures from 20 to 110 GPa (128 molecules, 0.5 ns trajectories). 
  Blue: Experimental results converted from imaginary part of the dielectric function $\epsilon_2(\omega)$ of H${}_2$O at 295 K and various pressures obtained from a Kramers-Kronig analysis of experimental reflectivity spectra~\cite{goncharov1996compression}, according to the relationship $n(\omega)\alpha(\omega)=4\pi\omega\epsilon_2(\omega)$.
  Dashed lines denote oscillator fit since there are no data in the range of strong diamond absorption (1800 to 2400 cm${}^{-1}$).
  The curves are offset in the vertical direction for clarity.
    \label{fig:ir-ice}}
\end{figure}

Our results in Fig.~\ref{fig:ir-ice} show a dramatic change with pressure of the product $\alpha(\omega) n(\omega)$. 
We also report the same quantity obtained from reflectance measurements of pressurized ice in a diamond anvil cell~\cite{goncharov1996compression}. 
Experimental data are not available for $\omega\lesssim 800$ cm${}^{-1}$, and for 1800 cm${}^{-1} < \omega < $2400 cm${}^{-1}$.
The theoretical curves are displayed at approximately 10 GPa higher pressure than the experimental curves to empirically correct for the missing nuclear quantum effects in  the simulation and the inaccuracy of the adopted functional approximation.                  

Taking into account the limitations of theory and experiment, the experimental trend is reproduced nicely by the DP+DW simulations. 
At the lowest pressures, molecular vibration features can be discerned, such as the H stretching band at about 3000 cm${}^{-1}$. 
These modes soften dramatically and broaden with pressure, indicating a progressive weakening of the covalent O-H bonds. 
In the simulation, the spectral changes are quite abrupt at $\sim$70 GPa, suggesting that this should be approximately the transition pressure to symmetric ice X. 
The same behavior is observed in experiment at $\sim$60 GPa. 
By further increasing the pressure, two strong features emerge, characteristic of ice X, that harden and sharpen with pressure. 
The higher frequency feature has dominant weight on H while the lower frequency feature has more O weight. 
Interestingly, there is no close correspondence between the IR spectrum and the power spectrum of atomic dynamics reported in the [SM], suggesting that extreme anharmonicity affects the IR spectrum, as pointed out in Ref.~\cite{bernasconi1998ab}. 

Quantum effects in the dynamics were ignored in our calculations. 
These effects are small but non-negligible in liquid water and should be more pronounced near the ice VII to X transition in view of the relatively large isotope
effect on the transition pressure, which is $\sim$10 GPa higher for D${}_2$O than for H${}_2$O~\cite{goncharov1996compression}, 
suggesting that the transition is facilitated by proton tunneling.
The calculation of dynamic quantum correlation functions is a major challenge for statistical simulations. 
Quantum IR correlations have been calculated recently for liquid water using approximate methods like  ring polymer~\cite{marsalek2017quantum} and centroid~\cite{marsalek2017quantum,medders2015infrared} MD, indicating that quantum corrections tend to red shift the classical spectral features. 
It would be extremely interesting to study how quantum corrections affect the dielectric properties in the ice VII, VII', and X transition sequence.
Further studies of these issues will be facilitated by methods like DPMD and DW, which improve significantly the statistical quality of {\it ab initio} simulations, as they are orders of magnitude faster than DFT methods and scale linearly with system size.
Quantitative cost comparisons between direct AIMD and DP/DW simulations are reported in Figs. S2 and S3 in the [SM]. 

In summary, DW is a useful tool to parametrize the dependence of the polarization on the atomic environment. 
The approach can be naturally extended~\cite{grace2020raman} to the environmental dependence of the polarizability tensor  $P_{\delta\sigma}=\frac{\partial M_\delta}{\partial E_\sigma}$ ($M_\delta$ and $E_\sigma$ are Cartesian components of the polarization and of the electric field),
allowing us to compute Raman~\cite{wan2013raman,grace2020raman} and sum frequency generation spectra~\cite{Mukamel1998Principles,Nagata2010Vibrational}. 
Access to the concerted evolution of atomic coordinates and polarization in simulations of large systems over long  time scales should also open the way to studies of ferroelectric phase transitions with MD simulations of {\it ab initio} quality, rather than relying on effective Hamiltonian models~\cite{resta2007theory,srinivasan2003pbtio3,david1998first}.     
Finally, related symmetry preserving DNN schemes have been considered in Refs.~\cite{Zaheer2017deepsets,battaglia2018relational,esteves2018learning,han2019universal} and we defer to future work a discussion of the mathematical and machine learning aspects of the DP/DW models.

\begin{acknowledgements}
The work of L. Z., X. W., W. E, and R.C. was conducted at the Center ``Chemistry in Solution and at Interfaces'' (CSI) funded by the DOE Award DE-SC001934. 
The work of L. Z and W. E was partially supported by a gift from iFlytek to Princeton University and by ONR grant N00014-13-1-0338.
The work of H.W. was supported by the NSFC under grant 11871110, and the National Key Research and Development Program of China under grants 2016YFB0201200 and 2016YFB0201203.
We used resources of the National Energy Research Scientific Computing Center (DoE Contract No. DE-AC02-05cH11231). 
We are also grateful for computing time at the Terascale Infrastructure for Groundbreaking Research in Science and Engineering (TIGRESS) of Princeton University.
\end{acknowledgements}


\begin{thebibliography}{72}%
\makeatletter
\providecommand \@ifxundefined [1]{%
 \@ifx{#1\undefined}
}%
\providecommand \@ifnum [1]{%
 \ifnum #1\expandafter \@firstoftwo
 \else \expandafter \@secondoftwo
 \fi
}%
\providecommand \@ifx [1]{%
 \ifx #1\expandafter \@firstoftwo
 \else \expandafter \@secondoftwo
 \fi
}%
\providecommand \natexlab [1]{#1}%
\providecommand \enquote  [1]{``#1''}%
\providecommand \bibnamefont  [1]{#1}%
\providecommand \bibfnamefont [1]{#1}%
\providecommand \citenamefont [1]{#1}%
\providecommand \href@noop [0]{\@secondoftwo}%
\providecommand \href [0]{\begingroup \@sanitize@url \@href}%
\providecommand \@href[1]{\@@startlink{#1}\@@href}%
\providecommand \@@href[1]{\endgroup#1\@@endlink}%
\providecommand \@sanitize@url [0]{\catcode `\\12\catcode `\$12\catcode
  `\&12\catcode `\#12\catcode `\^12\catcode `\_12\catcode `\%12\relax}%
\providecommand \@@startlink[1]{}%
\providecommand \@@endlink[0]{}%
\providecommand \url  [0]{\begingroup\@sanitize@url \@url }%
\providecommand \@url [1]{\endgroup\@href {#1}{\urlprefix }}%
\providecommand \urlprefix  [0]{URL }%
\providecommand \Eprint [0]{\href }%
\providecommand \doibase [0]{http://dx.doi.org/}%
\providecommand \selectlanguage [0]{\@gobble}%
\providecommand \bibinfo  [0]{\@secondoftwo}%
\providecommand \bibfield  [0]{\@secondoftwo}%
\providecommand \translation [1]{[#1]}%
\providecommand \BibitemOpen [0]{}%
\providecommand \bibitemStop [0]{}%
\providecommand \bibitemNoStop [0]{.\EOS\space}%
\providecommand \EOS [0]{\spacefactor3000\relax}%
\providecommand \BibitemShut  [1]{\csname bibitem#1\endcsname}%
\let\auto@bib@innerbib\@empty
\bibitem [{\citenamefont {Behler}\ and\ \citenamefont
  {Parrinello}(2007)}]{behler2007generalized}%
  \BibitemOpen
  \bibfield  {author} {\bibinfo {author} {\bibfnamefont {J.}~\bibnamefont
  {Behler}}\ and\ \bibinfo {author} {\bibfnamefont {M.}~\bibnamefont
  {Parrinello}},\ }\href@noop {} {\bibfield  {journal} {\bibinfo  {journal}
  {Physical Review Letters}\ }\textbf {\bibinfo {volume} {98}},\ \bibinfo
  {pages} {146401} (\bibinfo {year} {2007})}\BibitemShut {NoStop}%
\bibitem [{\citenamefont {Bart{\'o}k}\ \emph {et~al.}(2010)\citenamefont
  {Bart{\'o}k}, \citenamefont {Payne}, \citenamefont {Kondor},\ and\
  \citenamefont {Cs{\'a}nyi}}]{bartok2010gaussian}%
  \BibitemOpen
  \bibfield  {author} {\bibinfo {author} {\bibfnamefont {A.~P.}\ \bibnamefont
  {Bart{\'o}k}}, \bibinfo {author} {\bibfnamefont {M.~C.}\ \bibnamefont
  {Payne}}, \bibinfo {author} {\bibfnamefont {R.}~\bibnamefont {Kondor}}, \
  and\ \bibinfo {author} {\bibfnamefont {G.}~\bibnamefont {Cs{\'a}nyi}},\
  }\href@noop {} {\bibfield  {journal} {\bibinfo  {journal} {Physical Review
  Letters}\ }\textbf {\bibinfo {volume} {104}},\ \bibinfo {pages} {136403}
  (\bibinfo {year} {2010})}\BibitemShut {NoStop}%
\bibitem [{\citenamefont {Rupp}\ \emph {et~al.}(2012)\citenamefont {Rupp},
  \citenamefont {Tkatchenko}, \citenamefont {M{\"u}ller},\ and\ \citenamefont
  {VonLilienfeld}}]{rupp2012fast}%
  \BibitemOpen
  \bibfield  {author} {\bibinfo {author} {\bibfnamefont {M.}~\bibnamefont
  {Rupp}}, \bibinfo {author} {\bibfnamefont {A.}~\bibnamefont {Tkatchenko}},
  \bibinfo {author} {\bibfnamefont {K.-R.}\ \bibnamefont {M{\"u}ller}}, \ and\
  \bibinfo {author} {\bibfnamefont {O.~A.}\ \bibnamefont {VonLilienfeld}},\
  }\href@noop {} {\bibfield  {journal} {\bibinfo  {journal} {Physical Review
  Letters}\ }\textbf {\bibinfo {volume} {108}},\ \bibinfo {pages} {058301}
  (\bibinfo {year} {2012})}\BibitemShut {NoStop}%
\bibitem [{\citenamefont {Montavon}\ \emph {et~al.}(2013)\citenamefont
  {Montavon}, \citenamefont {Rupp}, \citenamefont {Gobre}, \citenamefont
  {Vazquez-Mayagoitia}, \citenamefont {Hansen}, \citenamefont {Tkatchenko},
  \citenamefont {M{\"u}ller},\ and\ \citenamefont
  {Von~Lilienfeld}}]{montavon2013machine}%
  \BibitemOpen
  \bibfield  {author} {\bibinfo {author} {\bibfnamefont {G.}~\bibnamefont
  {Montavon}}, \bibinfo {author} {\bibfnamefont {M.}~\bibnamefont {Rupp}},
  \bibinfo {author} {\bibfnamefont {V.}~\bibnamefont {Gobre}}, \bibinfo
  {author} {\bibfnamefont {A.}~\bibnamefont {Vazquez-Mayagoitia}}, \bibinfo
  {author} {\bibfnamefont {K.}~\bibnamefont {Hansen}}, \bibinfo {author}
  {\bibfnamefont {A.}~\bibnamefont {Tkatchenko}}, \bibinfo {author}
  {\bibfnamefont {K.-R.}\ \bibnamefont {M{\"u}ller}}, \ and\ \bibinfo {author}
  {\bibfnamefont {O.~A.}\ \bibnamefont {Von~Lilienfeld}},\ }\href@noop {}
  {\bibfield  {journal} {\bibinfo  {journal} {New Journal of Physics}\ }\textbf
  {\bibinfo {volume} {15}},\ \bibinfo {pages} {095003} (\bibinfo {year}
  {2013})}\BibitemShut {NoStop}%
\bibitem [{\citenamefont {Botu}\ \emph {et~al.}(2016)\citenamefont {Botu},
  \citenamefont {Batra}, \citenamefont {Chapman},\ and\ \citenamefont
  {Ramprasad}}]{botu2016machine}%
  \BibitemOpen
  \bibfield  {author} {\bibinfo {author} {\bibfnamefont {V.}~\bibnamefont
  {Botu}}, \bibinfo {author} {\bibfnamefont {R.}~\bibnamefont {Batra}},
  \bibinfo {author} {\bibfnamefont {J.}~\bibnamefont {Chapman}}, \ and\
  \bibinfo {author} {\bibfnamefont {R.}~\bibnamefont {Ramprasad}},\ }\href@noop
  {} {\bibfield  {journal} {\bibinfo  {journal} {The Journal of Physical
  Chemistry C}\ }\textbf {\bibinfo {volume} {121}},\ \bibinfo {pages} {511}
  (\bibinfo {year} {2016})}\BibitemShut {NoStop}%
\bibitem [{\citenamefont {Chmiela}\ \emph {et~al.}(2017)\citenamefont
  {Chmiela}, \citenamefont {Tkatchenko}, \citenamefont {Sauceda}, \citenamefont
  {Poltavsky}, \citenamefont {Sch{\"u}tt},\ and\ \citenamefont
  {M{\"u}ller}}]{chmiela2017machine}%
  \BibitemOpen
  \bibfield  {author} {\bibinfo {author} {\bibfnamefont {S.}~\bibnamefont
  {Chmiela}}, \bibinfo {author} {\bibfnamefont {A.}~\bibnamefont {Tkatchenko}},
  \bibinfo {author} {\bibfnamefont {H.~E.}\ \bibnamefont {Sauceda}}, \bibinfo
  {author} {\bibfnamefont {I.}~\bibnamefont {Poltavsky}}, \bibinfo {author}
  {\bibfnamefont {K.~T.}\ \bibnamefont {Sch{\"u}tt}}, \ and\ \bibinfo {author}
  {\bibfnamefont {K.-R.}\ \bibnamefont {M{\"u}ller}},\ }\href@noop {}
  {\bibfield  {journal} {\bibinfo  {journal} {Science Advances}\ }\textbf
  {\bibinfo {volume} {3}},\ \bibinfo {pages} {e1603015} (\bibinfo {year}
  {2017})}\BibitemShut {NoStop}%
\bibitem [{\citenamefont {Sch{\"u}tt}\ \emph {et~al.}(2017)\citenamefont
  {Sch{\"u}tt}, \citenamefont {Kindermans}, \citenamefont {Felix},
  \citenamefont {Chmiela}, \citenamefont {Tkatchenko},\ and\ \citenamefont
  {M{\"u}ller}}]{schutt2017schnet}%
  \BibitemOpen
  \bibfield  {author} {\bibinfo {author} {\bibfnamefont {K.}~\bibnamefont
  {Sch{\"u}tt}}, \bibinfo {author} {\bibfnamefont {P.-J.}\ \bibnamefont
  {Kindermans}}, \bibinfo {author} {\bibfnamefont {H.~E.~S.}\ \bibnamefont
  {Felix}}, \bibinfo {author} {\bibfnamefont {S.}~\bibnamefont {Chmiela}},
  \bibinfo {author} {\bibfnamefont {A.}~\bibnamefont {Tkatchenko}}, \ and\
  \bibinfo {author} {\bibfnamefont {K.-R.}\ \bibnamefont {M{\"u}ller}},\ }in\
  \href@noop {} {\emph {\bibinfo {booktitle} {Advances in Neural Information
  Processing Systems}}}\ (\bibinfo {year} {2017})\ pp.\ \bibinfo {pages}
  {992--1002}\BibitemShut {NoStop}%
\bibitem [{\citenamefont {Smith}\ \emph {et~al.}(2017)\citenamefont {Smith},
  \citenamefont {Isayev},\ and\ \citenamefont {Roitberg}}]{smith2017ani}%
  \BibitemOpen
  \bibfield  {author} {\bibinfo {author} {\bibfnamefont {J.~S.}\ \bibnamefont
  {Smith}}, \bibinfo {author} {\bibfnamefont {O.}~\bibnamefont {Isayev}}, \
  and\ \bibinfo {author} {\bibfnamefont {A.~E.}\ \bibnamefont {Roitberg}},\
  }\href@noop {} {\bibfield  {journal} {\bibinfo  {journal} {Chemical Science}\
  }\textbf {\bibinfo {volume} {8}},\ \bibinfo {pages} {3192} (\bibinfo {year}
  {2017})}\BibitemShut {NoStop}%
\bibitem [{\citenamefont {Han}\ \emph {et~al.}(2018)\citenamefont {Han},
  \citenamefont {Zhang}, \citenamefont {Car},\ and\ \citenamefont
  {E}}]{han2017deep}%
  \BibitemOpen
  \bibfield  {author} {\bibinfo {author} {\bibfnamefont {J.}~\bibnamefont
  {Han}}, \bibinfo {author} {\bibfnamefont {L.}~\bibnamefont {Zhang}}, \bibinfo
  {author} {\bibfnamefont {R.}~\bibnamefont {Car}}, \ and\ \bibinfo {author}
  {\bibfnamefont {W.}~\bibnamefont {E}},\ }\href@noop {} {\bibfield  {journal}
  {\bibinfo  {journal} {Communications in Computational Physics}\ }\textbf
  {\bibinfo {volume} {23}},\ \bibinfo {pages} {629} (\bibinfo {year}
  {2018})}\BibitemShut {NoStop}%
\bibitem [{\citenamefont {Zhang}\ \emph
  {et~al.}(2018{\natexlab{a}})\citenamefont {Zhang}, \citenamefont {Han},
  \citenamefont {Wang}, \citenamefont {Car},\ and\ \citenamefont
  {E}}]{zhang2018deep}%
  \BibitemOpen
  \bibfield  {author} {\bibinfo {author} {\bibfnamefont {L.}~\bibnamefont
  {Zhang}}, \bibinfo {author} {\bibfnamefont {J.}~\bibnamefont {Han}}, \bibinfo
  {author} {\bibfnamefont {H.}~\bibnamefont {Wang}}, \bibinfo {author}
  {\bibfnamefont {R.}~\bibnamefont {Car}}, \ and\ \bibinfo {author}
  {\bibfnamefont {W.}~\bibnamefont {E}},\ }\href@noop {} {\bibfield  {journal}
  {\bibinfo  {journal} {Physical Review Letters}\ }\textbf {\bibinfo {volume}
  {120}},\ \bibinfo {pages} {143001} (\bibinfo {year}
  {2018}{\natexlab{a}})}\BibitemShut {NoStop}%
\bibitem [{\citenamefont {Zhang}\ \emph
  {et~al.}(2018{\natexlab{b}})\citenamefont {Zhang}, \citenamefont {Han},
  \citenamefont {Wang}, \citenamefont {Saidi}, \citenamefont {Car},\ and\
  \citenamefont {E}}]{zhang2018end}%
  \BibitemOpen
  \bibfield  {author} {\bibinfo {author} {\bibfnamefont {L.}~\bibnamefont
  {Zhang}}, \bibinfo {author} {\bibfnamefont {J.}~\bibnamefont {Han}}, \bibinfo
  {author} {\bibfnamefont {H.}~\bibnamefont {Wang}}, \bibinfo {author}
  {\bibfnamefont {W.}~\bibnamefont {Saidi}}, \bibinfo {author} {\bibfnamefont
  {R.}~\bibnamefont {Car}}, \ and\ \bibinfo {author} {\bibfnamefont
  {W.}~\bibnamefont {E}},\ }in\ \href@noop {} {\emph {\bibinfo {booktitle}
  {Advances in Neural Information Processing Systems 31}}},\ \bibinfo {editor}
  {edited by\ \bibinfo {editor} {\bibfnamefont {S.}~\bibnamefont {Bengio}},
  \bibinfo {editor} {\bibfnamefont {H.}~\bibnamefont {Wallach}}, \bibinfo
  {editor} {\bibfnamefont {H.}~\bibnamefont {Larochelle}}, \bibinfo {editor}
  {\bibfnamefont {K.}~\bibnamefont {Grauman}}, \bibinfo {editor} {\bibfnamefont
  {N.}~\bibnamefont {Cesa-Bianchi}}, \ and\ \bibinfo {editor} {\bibfnamefont
  {R.}~\bibnamefont {Garnett}}}\ (\bibinfo  {publisher} {Curran Associates,
  Inc.},\ \bibinfo {year} {2018})\ pp.\ \bibinfo {pages}
  {4441--4451}\BibitemShut {NoStop}%
\bibitem [{\citenamefont {Kohn}\ and\ \citenamefont
  {Sham}(1965)}]{kohn1965self}%
  \BibitemOpen
  \bibfield  {author} {\bibinfo {author} {\bibfnamefont {W.}~\bibnamefont
  {Kohn}}\ and\ \bibinfo {author} {\bibfnamefont {L.~J.}\ \bibnamefont
  {Sham}},\ }\href@noop {} {\bibfield  {journal} {\bibinfo  {journal} {Physical
  Review}\ }\textbf {\bibinfo {volume} {140}},\ \bibinfo {pages} {A1133}
  (\bibinfo {year} {1965})}\BibitemShut {NoStop}%
\bibitem [{\citenamefont {Car}\ and\ \citenamefont
  {Parrinello}(1985)}]{car1985unified}%
  \BibitemOpen
  \bibfield  {author} {\bibinfo {author} {\bibfnamefont {R.}~\bibnamefont
  {Car}}\ and\ \bibinfo {author} {\bibfnamefont {M.}~\bibnamefont
  {Parrinello}},\ }\href@noop {} {\bibfield  {journal} {\bibinfo  {journal}
  {Physical Review Letters}\ }\textbf {\bibinfo {volume} {55}},\ \bibinfo
  {pages} {2471} (\bibinfo {year} {1985})}\BibitemShut {NoStop}%
\bibitem [{\citenamefont {Marx}\ and\ \citenamefont
  {Hutter}(2009)}]{marx2009ab}%
  \BibitemOpen
  \bibfield  {author} {\bibinfo {author} {\bibfnamefont {D.}~\bibnamefont
  {Marx}}\ and\ \bibinfo {author} {\bibfnamefont {J.}~\bibnamefont {Hutter}},\
  }\href@noop {} {\emph {\bibinfo {title} {Ab initio molecular dynamics: basic
  theory and advanced methods}}}\ (\bibinfo  {publisher} {Cambridge University
  Press},\ \bibinfo {year} {2009})\BibitemShut {NoStop}%
\bibitem [{\citenamefont {Bonati}\ and\ \citenamefont
  {Parrinello}(2018)}]{bonati2018silicon}%
  \BibitemOpen
  \bibfield  {author} {\bibinfo {author} {\bibfnamefont {L.}~\bibnamefont
  {Bonati}}\ and\ \bibinfo {author} {\bibfnamefont {M.}~\bibnamefont
  {Parrinello}},\ }\href@noop {} {\bibfield  {journal} {\bibinfo  {journal}
  {Physical Review Letters}\ }\textbf {\bibinfo {volume} {121}},\ \bibinfo
  {pages} {265701} (\bibinfo {year} {2018})}\BibitemShut {NoStop}%
\bibitem [{\citenamefont {Gastegger}\ \emph {et~al.}(2017)\citenamefont
  {Gastegger}, \citenamefont {Behler},\ and\ \citenamefont
  {Marquetand}}]{gastegger2017machine}%
  \BibitemOpen
  \bibfield  {author} {\bibinfo {author} {\bibfnamefont {M.}~\bibnamefont
  {Gastegger}}, \bibinfo {author} {\bibfnamefont {J.}~\bibnamefont {Behler}}, \
  and\ \bibinfo {author} {\bibfnamefont {P.}~\bibnamefont {Marquetand}},\
  }\href@noop {} {\bibfield  {journal} {\bibinfo  {journal} {Chemical Science}\
  }\textbf {\bibinfo {volume} {8}},\ \bibinfo {pages} {6924} (\bibinfo {year}
  {2017})}\BibitemShut {NoStop}%
\bibitem [{\citenamefont {Grisafi}\ \emph
  {et~al.}(2018{\natexlab{a}})\citenamefont {Grisafi}, \citenamefont {Wilkins},
  \citenamefont {Cs{\'a}nyi},\ and\ \citenamefont
  {Ceriotti}}]{grisafi2018symmetry}%
  \BibitemOpen
  \bibfield  {author} {\bibinfo {author} {\bibfnamefont {A.}~\bibnamefont
  {Grisafi}}, \bibinfo {author} {\bibfnamefont {D.~M.}\ \bibnamefont
  {Wilkins}}, \bibinfo {author} {\bibfnamefont {G.}~\bibnamefont {Cs{\'a}nyi}},
  \ and\ \bibinfo {author} {\bibfnamefont {M.}~\bibnamefont {Ceriotti}},\
  }\href@noop {} {\bibfield  {journal} {\bibinfo  {journal} {Physical Review
  Letters}\ }\textbf {\bibinfo {volume} {120}},\ \bibinfo {pages} {036002}
  (\bibinfo {year} {2018}{\natexlab{a}})}\BibitemShut {NoStop}%
\bibitem [{\citenamefont {Grisafi}\ \emph
  {et~al.}(2018{\natexlab{b}})\citenamefont {Grisafi}, \citenamefont
  {Fabrizio}, \citenamefont {Meyer}, \citenamefont {Wilkins}, \citenamefont
  {Corminboeuf},\ and\ \citenamefont {Ceriotti}}]{grisafi2018transferable}%
  \BibitemOpen
  \bibfield  {author} {\bibinfo {author} {\bibfnamefont {A.}~\bibnamefont
  {Grisafi}}, \bibinfo {author} {\bibfnamefont {A.}~\bibnamefont {Fabrizio}},
  \bibinfo {author} {\bibfnamefont {B.}~\bibnamefont {Meyer}}, \bibinfo
  {author} {\bibfnamefont {D.~M.}\ \bibnamefont {Wilkins}}, \bibinfo {author}
  {\bibfnamefont {C.}~\bibnamefont {Corminboeuf}}, \ and\ \bibinfo {author}
  {\bibfnamefont {M.}~\bibnamefont {Ceriotti}},\ }\href@noop {} {\bibfield
  {journal} {\bibinfo  {journal} {ACS Central Science}\ }\textbf {\bibinfo
  {volume} {5}},\ \bibinfo {pages} {57} (\bibinfo {year}
  {2018}{\natexlab{b}})}\BibitemShut {NoStop}%
\bibitem [{\citenamefont {Wilkins}\ \emph {et~al.}(2019)\citenamefont
  {Wilkins}, \citenamefont {Grisafi}, \citenamefont {Yang}, \citenamefont
  {Lao}, \citenamefont {DiStasio},\ and\ \citenamefont
  {Ceriotti}}]{wilkins2019accurate}%
  \BibitemOpen
  \bibfield  {author} {\bibinfo {author} {\bibfnamefont {D.~M.}\ \bibnamefont
  {Wilkins}}, \bibinfo {author} {\bibfnamefont {A.}~\bibnamefont {Grisafi}},
  \bibinfo {author} {\bibfnamefont {Y.}~\bibnamefont {Yang}}, \bibinfo {author}
  {\bibfnamefont {K.~U.}\ \bibnamefont {Lao}}, \bibinfo {author} {\bibfnamefont
  {R.~A.}\ \bibnamefont {DiStasio}}, \ and\ \bibinfo {author} {\bibfnamefont
  {M.}~\bibnamefont {Ceriotti}},\ }\href@noop {} {\bibfield  {journal}
  {\bibinfo  {journal} {Proceedings of the National Academy of Sciences}\ ,\
  \bibinfo {pages} {201816132}} (\bibinfo {year} {2019})}\BibitemShut {NoStop}%
\bibitem [{\citenamefont {Chandrasekaran}\ \emph {et~al.}(2019)\citenamefont
  {Chandrasekaran}, \citenamefont {Kamal}, \citenamefont {Batra}, \citenamefont
  {Kim}, \citenamefont {Chen},\ and\ \citenamefont
  {Ramprasad}}]{chandrasekaran2019solving}%
  \BibitemOpen
  \bibfield  {author} {\bibinfo {author} {\bibfnamefont {A.}~\bibnamefont
  {Chandrasekaran}}, \bibinfo {author} {\bibfnamefont {D.}~\bibnamefont
  {Kamal}}, \bibinfo {author} {\bibfnamefont {R.}~\bibnamefont {Batra}},
  \bibinfo {author} {\bibfnamefont {C.}~\bibnamefont {Kim}}, \bibinfo {author}
  {\bibfnamefont {L.}~\bibnamefont {Chen}}, \ and\ \bibinfo {author}
  {\bibfnamefont {R.}~\bibnamefont {Ramprasad}},\ }\href@noop {} {\bibfield
  {journal} {\bibinfo  {journal} {NPJ Computational Materials}\ }\textbf
  {\bibinfo {volume} {5}},\ \bibinfo {pages} {22} (\bibinfo {year}
  {2019})}\BibitemShut {NoStop}%
\bibitem [{\citenamefont {Zepeda-N{\'u}{\~n}ez}\ \emph
  {et~al.}(2019)\citenamefont {Zepeda-N{\'u}{\~n}ez}, \citenamefont {Chen},
  \citenamefont {Zhang}, \citenamefont {Jia}, \citenamefont {Zhang},\ and\
  \citenamefont {Lin}}]{zepeda2019deep}%
  \BibitemOpen
  \bibfield  {author} {\bibinfo {author} {\bibfnamefont {L.}~\bibnamefont
  {Zepeda-N{\'u}{\~n}ez}}, \bibinfo {author} {\bibfnamefont {Y.}~\bibnamefont
  {Chen}}, \bibinfo {author} {\bibfnamefont {J.}~\bibnamefont {Zhang}},
  \bibinfo {author} {\bibfnamefont {W.}~\bibnamefont {Jia}}, \bibinfo {author}
  {\bibfnamefont {L.}~\bibnamefont {Zhang}}, \ and\ \bibinfo {author}
  {\bibfnamefont {L.}~\bibnamefont {Lin}},\ }\href@noop {} {\bibfield
  {journal} {\bibinfo  {journal} {arXiv preprint arXiv:1912.00775}\ } (\bibinfo
  {year} {2019})}\BibitemShut {NoStop}%
\bibitem [{\citenamefont {Raimbault}\ \emph {et~al.}(2019)\citenamefont
  {Raimbault}, \citenamefont {Grisafi}, \citenamefont {Ceriotti},\ and\
  \citenamefont {Rossi}}]{raimbault2019using}%
  \BibitemOpen
  \bibfield  {author} {\bibinfo {author} {\bibfnamefont {N.}~\bibnamefont
  {Raimbault}}, \bibinfo {author} {\bibfnamefont {A.}~\bibnamefont {Grisafi}},
  \bibinfo {author} {\bibfnamefont {M.}~\bibnamefont {Ceriotti}}, \ and\
  \bibinfo {author} {\bibfnamefont {M.}~\bibnamefont {Rossi}},\ }\href@noop {}
  {\bibfield  {journal} {\bibinfo  {journal} {New Journal of Physics}\ }\textbf
  {\bibinfo {volume} {21}},\ \bibinfo {pages} {105001} (\bibinfo {year}
  {2019})}\BibitemShut {NoStop}%
\bibitem [{\citenamefont {Kapil}\ \emph {et~al.}(2019)\citenamefont {Kapil},
  \citenamefont {Wilkins}, \citenamefont {Lan},\ and\ \citenamefont
  {Ceriotti}}]{kapil2019inexpensive}%
  \BibitemOpen
  \bibfield  {author} {\bibinfo {author} {\bibfnamefont {V.}~\bibnamefont
  {Kapil}}, \bibinfo {author} {\bibfnamefont {D.~M.}\ \bibnamefont {Wilkins}},
  \bibinfo {author} {\bibfnamefont {J.}~\bibnamefont {Lan}}, \ and\ \bibinfo
  {author} {\bibfnamefont {M.}~\bibnamefont {Ceriotti}},\ }\href@noop {}
  {\bibfield  {journal} {\bibinfo  {journal} {arXiv preprint arXiv:1912.03189}\
  } (\bibinfo {year} {2019})}\BibitemShut {NoStop}%
\bibitem [{\citenamefont {Sharma}\ \emph {et~al.}(2005)\citenamefont {Sharma},
  \citenamefont {Resta},\ and\ \citenamefont {Car}}]{sharma2005intermolecular}%
  \BibitemOpen
  \bibfield  {author} {\bibinfo {author} {\bibfnamefont {M.}~\bibnamefont
  {Sharma}}, \bibinfo {author} {\bibfnamefont {R.}~\bibnamefont {Resta}}, \
  and\ \bibinfo {author} {\bibfnamefont {R.}~\bibnamefont {Car}},\ }\href@noop
  {} {\bibfield  {journal} {\bibinfo  {journal} {Physical Review Letters}\
  }\textbf {\bibinfo {volume} {95}},\ \bibinfo {pages} {187401} (\bibinfo
  {year} {2005})}\BibitemShut {NoStop}%
\bibitem [{\citenamefont {Putrino}\ \emph {et~al.}(2000)\citenamefont
  {Putrino}, \citenamefont {Sebastiani},\ and\ \citenamefont
  {Parrinello}}]{putrino2000generalized}%
  \BibitemOpen
  \bibfield  {author} {\bibinfo {author} {\bibfnamefont {A.}~\bibnamefont
  {Putrino}}, \bibinfo {author} {\bibfnamefont {D.}~\bibnamefont {Sebastiani}},
  \ and\ \bibinfo {author} {\bibfnamefont {M.}~\bibnamefont {Parrinello}},\
  }\href@noop {} {\bibfield  {journal} {\bibinfo  {journal} {The Journal of
  Chemical Physics}\ }\textbf {\bibinfo {volume} {113}},\ \bibinfo {pages}
  {7102} (\bibinfo {year} {2000})}\BibitemShut {NoStop}%
\bibitem [{\citenamefont {Wan}\ \emph {et~al.}(2013)\citenamefont {Wan},
  \citenamefont {Spanu}, \citenamefont {Galli},\ and\ \citenamefont
  {Gygi}}]{wan2013raman}%
  \BibitemOpen
  \bibfield  {author} {\bibinfo {author} {\bibfnamefont {Q.}~\bibnamefont
  {Wan}}, \bibinfo {author} {\bibfnamefont {L.}~\bibnamefont {Spanu}}, \bibinfo
  {author} {\bibfnamefont {G.~A.}\ \bibnamefont {Galli}}, \ and\ \bibinfo
  {author} {\bibfnamefont {F.}~\bibnamefont {Gygi}},\ }\href@noop {} {\bibfield
   {journal} {\bibinfo  {journal} {Journal of Chemical Theory and Computation}\
  }\textbf {\bibinfo {volume} {9}},\ \bibinfo {pages} {4124} (\bibinfo {year}
  {2013})}\BibitemShut {NoStop}%
\bibitem [{\citenamefont {Wan}\ and\ \citenamefont
  {Galli}(2015)}]{wan2015first}%
  \BibitemOpen
  \bibfield  {author} {\bibinfo {author} {\bibfnamefont {Q.}~\bibnamefont
  {Wan}}\ and\ \bibinfo {author} {\bibfnamefont {G.}~\bibnamefont {Galli}},\
  }\href@noop {} {\bibfield  {journal} {\bibinfo  {journal} {Physical Review
  Letters}\ }\textbf {\bibinfo {volume} {115}},\ \bibinfo {pages} {246404}
  (\bibinfo {year} {2015})}\BibitemShut {NoStop}%
\bibitem [{\citenamefont {Rozsa}\ \emph {et~al.}(2018)\citenamefont {Rozsa},
  \citenamefont {Pan}, \citenamefont {Giberti},\ and\ \citenamefont
  {Galli}}]{rozsa2018ab}%
  \BibitemOpen
  \bibfield  {author} {\bibinfo {author} {\bibfnamefont {V.}~\bibnamefont
  {Rozsa}}, \bibinfo {author} {\bibfnamefont {D.}~\bibnamefont {Pan}}, \bibinfo
  {author} {\bibfnamefont {F.}~\bibnamefont {Giberti}}, \ and\ \bibinfo
  {author} {\bibfnamefont {G.}~\bibnamefont {Galli}},\ }\href@noop {}
  {\bibfield  {journal} {\bibinfo  {journal} {Proceedings of the National
  Academy of Sciences}\ }\textbf {\bibinfo {volume} {115}},\ \bibinfo {pages}
  {6952} (\bibinfo {year} {2018})}\BibitemShut {NoStop}%
\bibitem [{\citenamefont {Sun}\ \emph {et~al.}(2015)\citenamefont {Sun},
  \citenamefont {Clark}, \citenamefont {Torquato},\ and\ \citenamefont
  {Car}}]{sun2015phase}%
  \BibitemOpen
  \bibfield  {author} {\bibinfo {author} {\bibfnamefont {J.}~\bibnamefont
  {Sun}}, \bibinfo {author} {\bibfnamefont {B.~K.}\ \bibnamefont {Clark}},
  \bibinfo {author} {\bibfnamefont {S.}~\bibnamefont {Torquato}}, \ and\
  \bibinfo {author} {\bibfnamefont {R.}~\bibnamefont {Car}},\ }\href@noop {}
  {\bibfield  {journal} {\bibinfo  {journal} {Nature communications}\ }\textbf
  {\bibinfo {volume} {6}},\ \bibinfo {pages} {8156} (\bibinfo {year}
  {2015})}\BibitemShut {NoStop}%
\bibitem [{\citenamefont {Wood}\ and\ \citenamefont
  {Marzari}(2006)}]{wood2006dynamical}%
  \BibitemOpen
  \bibfield  {author} {\bibinfo {author} {\bibfnamefont {B.~C.}\ \bibnamefont
  {Wood}}\ and\ \bibinfo {author} {\bibfnamefont {N.}~\bibnamefont {Marzari}},\
  }\href@noop {} {\bibfield  {journal} {\bibinfo  {journal} {Physical review
  letters}\ }\textbf {\bibinfo {volume} {97}},\ \bibinfo {pages} {166401}
  (\bibinfo {year} {2006})}\BibitemShut {NoStop}%
\bibitem [{\citenamefont {Schwegler}\ \emph {et~al.}(2008)\citenamefont
  {Schwegler}, \citenamefont {Sharma}, \citenamefont {Gygi},\ and\
  \citenamefont {Galli}}]{schwegler2008melting}%
  \BibitemOpen
  \bibfield  {author} {\bibinfo {author} {\bibfnamefont {E.}~\bibnamefont
  {Schwegler}}, \bibinfo {author} {\bibfnamefont {M.}~\bibnamefont {Sharma}},
  \bibinfo {author} {\bibfnamefont {F.}~\bibnamefont {Gygi}}, \ and\ \bibinfo
  {author} {\bibfnamefont {G.}~\bibnamefont {Galli}},\ }\href@noop {}
  {\bibfield  {journal} {\bibinfo  {journal} {Proceedings of the National
  Academy of Sciences}\ }\textbf {\bibinfo {volume} {105}},\ \bibinfo {pages}
  {14779} (\bibinfo {year} {2008})}\BibitemShut {NoStop}%
\bibitem [{\citenamefont {Srinivasan}\ \emph {et~al.}(2003)\citenamefont
  {Srinivasan}, \citenamefont {Gebauer}, \citenamefont {Resta},\ and\
  \citenamefont {Car}}]{srinivasan2003pbtio3}%
  \BibitemOpen
  \bibfield  {author} {\bibinfo {author} {\bibfnamefont {V.}~\bibnamefont
  {Srinivasan}}, \bibinfo {author} {\bibfnamefont {R.}~\bibnamefont {Gebauer}},
  \bibinfo {author} {\bibfnamefont {R.}~\bibnamefont {Resta}}, \ and\ \bibinfo
  {author} {\bibfnamefont {R.}~\bibnamefont {Car}},\ }in\ \href@noop {} {\emph
  {\bibinfo {booktitle} {AIP Conference Proceedings}}},\ Vol.\ \bibinfo
  {volume} {677}\ (\bibinfo {organization} {AIP},\ \bibinfo {year} {2003})\
  pp.\ \bibinfo {pages} {168--175}\BibitemShut {NoStop}%
\bibitem [{\citenamefont {Fluri}\ \emph {et~al.}(2017)\citenamefont {Fluri},
  \citenamefont {Marcolongo}, \citenamefont {Roddatis}, \citenamefont {Wokaun},
  \citenamefont {Pergolesi}, \citenamefont {Marzari},\ and\ \citenamefont
  {Lippert}}]{fluri2017enhanced}%
  \BibitemOpen
  \bibfield  {author} {\bibinfo {author} {\bibfnamefont {A.}~\bibnamefont
  {Fluri}}, \bibinfo {author} {\bibfnamefont {A.}~\bibnamefont {Marcolongo}},
  \bibinfo {author} {\bibfnamefont {V.}~\bibnamefont {Roddatis}}, \bibinfo
  {author} {\bibfnamefont {A.}~\bibnamefont {Wokaun}}, \bibinfo {author}
  {\bibfnamefont {D.}~\bibnamefont {Pergolesi}}, \bibinfo {author}
  {\bibfnamefont {N.}~\bibnamefont {Marzari}}, \ and\ \bibinfo {author}
  {\bibfnamefont {T.}~\bibnamefont {Lippert}},\ }\href@noop {} {\bibfield
  {journal} {\bibinfo  {journal} {Advanced Science}\ }\textbf {\bibinfo
  {volume} {4}},\ \bibinfo {pages} {1700467} (\bibinfo {year}
  {2017})}\BibitemShut {NoStop}%
\bibitem [{\citenamefont {Liu}\ \emph {et~al.}(2015)\citenamefont {Liu},
  \citenamefont {Wang},\ and\ \citenamefont {Bowman}}]{liu2015quantum}%
  \BibitemOpen
  \bibfield  {author} {\bibinfo {author} {\bibfnamefont {H.}~\bibnamefont
  {Liu}}, \bibinfo {author} {\bibfnamefont {Y.}~\bibnamefont {Wang}}, \ and\
  \bibinfo {author} {\bibfnamefont {J.~M.}\ \bibnamefont {Bowman}},\
  }\href@noop {} {\bibfield  {journal} {\bibinfo  {journal} {The Journal of
  chemical physics}\ }\textbf {\bibinfo {volume} {142}},\ \bibinfo {pages}
  {194502} (\bibinfo {year} {2015})}\BibitemShut {NoStop}%
\bibitem [{\citenamefont {Marzari}\ and\ \citenamefont
  {Vanderbilt}(1997)}]{marzari1997maximally}%
  \BibitemOpen
  \bibfield  {author} {\bibinfo {author} {\bibfnamefont {N.}~\bibnamefont
  {Marzari}}\ and\ \bibinfo {author} {\bibfnamefont {D.}~\bibnamefont
  {Vanderbilt}},\ }\href@noop {} {\bibfield  {journal} {\bibinfo  {journal}
  {Physical Review B}\ }\textbf {\bibinfo {volume} {56}},\ \bibinfo {pages}
  {12847} (\bibinfo {year} {1997})}\BibitemShut {NoStop}%
\bibitem [{\citenamefont {Marzari}\ \emph {et~al.}(2012)\citenamefont
  {Marzari}, \citenamefont {Mostofi}, \citenamefont {Yates}, \citenamefont
  {Souza},\ and\ \citenamefont {Vanderbilt}}]{marzari2012maximally}%
  \BibitemOpen
  \bibfield  {author} {\bibinfo {author} {\bibfnamefont {N.}~\bibnamefont
  {Marzari}}, \bibinfo {author} {\bibfnamefont {A.~A.}\ \bibnamefont
  {Mostofi}}, \bibinfo {author} {\bibfnamefont {J.~R.}\ \bibnamefont {Yates}},
  \bibinfo {author} {\bibfnamefont {I.}~\bibnamefont {Souza}}, \ and\ \bibinfo
  {author} {\bibfnamefont {D.}~\bibnamefont {Vanderbilt}},\ }\href@noop {}
  {\bibfield  {journal} {\bibinfo  {journal} {Reviews of Modern Physics}\
  }\textbf {\bibinfo {volume} {84}},\ \bibinfo {pages} {1419} (\bibinfo {year}
  {2012})}\BibitemShut {NoStop}%
\bibitem [{\citenamefont {Resta}(1994)}]{Resta1994RMP}%
  \BibitemOpen
  \bibfield  {author} {\bibinfo {author} {\bibfnamefont {R.}~\bibnamefont
  {Resta}},\ }\href@noop {} {\bibfield  {journal} {\bibinfo  {journal} {Reviews
  of Modern Physics}\ }\textbf {\bibinfo {volume} {66}},\ \bibinfo {pages}
  {899} (\bibinfo {year} {1994})}\BibitemShut {NoStop}%
\bibitem [{\citenamefont {Kohn}(1996)}]{kohn1996density}%
  \BibitemOpen
  \bibfield  {author} {\bibinfo {author} {\bibfnamefont {W.}~\bibnamefont
  {Kohn}},\ }\href@noop {} {\bibfield  {journal} {\bibinfo  {journal} {Physical
  Review Letters}\ }\textbf {\bibinfo {volume} {76}},\ \bibinfo {pages} {3168}
  (\bibinfo {year} {1996})}\BibitemShut {NoStop}%
\bibitem [{\citenamefont {Prodan}\ and\ \citenamefont
  {Kohn}(2005)}]{prodan2005nearsightedness}%
  \BibitemOpen
  \bibfield  {author} {\bibinfo {author} {\bibfnamefont {E.}~\bibnamefont
  {Prodan}}\ and\ \bibinfo {author} {\bibfnamefont {W.}~\bibnamefont {Kohn}},\
  }\href@noop {} {\bibfield  {journal} {\bibinfo  {journal} {Proceedings of the
  National Academy of Sciences}\ }\textbf {\bibinfo {volume} {102}},\ \bibinfo
  {pages} {11635} (\bibinfo {year} {2005})}\BibitemShut {NoStop}%
\bibitem [{\citenamefont {Hernandez}\ and\ \citenamefont
  {Caracas}(2018)}]{hernandez2018proton}%
  \BibitemOpen
  \bibfield  {author} {\bibinfo {author} {\bibfnamefont {J.-A.}\ \bibnamefont
  {Hernandez}}\ and\ \bibinfo {author} {\bibfnamefont {R.}~\bibnamefont
  {Caracas}},\ }\href@noop {} {\bibfield  {journal} {\bibinfo  {journal} {The
  Journal of Chemical Physics}\ }\textbf {\bibinfo {volume} {148}},\ \bibinfo
  {pages} {214501} (\bibinfo {year} {2018})}\BibitemShut {NoStop}%
\bibitem [{\citenamefont {Bernal}\ and\ \citenamefont
  {Fowler}(1933)}]{bernal1933theory}%
  \BibitemOpen
  \bibfield  {author} {\bibinfo {author} {\bibfnamefont {J.~D.}\ \bibnamefont
  {Bernal}}\ and\ \bibinfo {author} {\bibfnamefont {R.~H.}\ \bibnamefont
  {Fowler}},\ }\href@noop {} {\bibfield  {journal} {\bibinfo  {journal} {The
  Journal of Chemical Physics}\ }\textbf {\bibinfo {volume} {1}},\ \bibinfo
  {pages} {515} (\bibinfo {year} {1933})}\BibitemShut {NoStop}%
\bibitem [{\citenamefont {Pauling}(1935)}]{pauling1935structure}%
  \BibitemOpen
  \bibfield  {author} {\bibinfo {author} {\bibfnamefont {L.}~\bibnamefont
  {Pauling}},\ }\href@noop {} {\bibfield  {journal} {\bibinfo  {journal}
  {Journal of the American Chemical Society}\ }\textbf {\bibinfo {volume}
  {57}},\ \bibinfo {pages} {2680} (\bibinfo {year} {1935})}\BibitemShut
  {NoStop}%
\bibitem [{\citenamefont {Millot}\ \emph {et~al.}(2019)\citenamefont {Millot},
  \citenamefont {Coppari}, \citenamefont {Rygg}, \citenamefont {Barrios},
  \citenamefont {Hamel}, \citenamefont {Swift},\ and\ \citenamefont
  {Eggert}}]{millot2019nanosecond}%
  \BibitemOpen
  \bibfield  {author} {\bibinfo {author} {\bibfnamefont {M.}~\bibnamefont
  {Millot}}, \bibinfo {author} {\bibfnamefont {F.}~\bibnamefont {Coppari}},
  \bibinfo {author} {\bibfnamefont {J.~R.}\ \bibnamefont {Rygg}}, \bibinfo
  {author} {\bibfnamefont {A.~C.}\ \bibnamefont {Barrios}}, \bibinfo {author}
  {\bibfnamefont {S.}~\bibnamefont {Hamel}}, \bibinfo {author} {\bibfnamefont
  {D.~C.}\ \bibnamefont {Swift}}, \ and\ \bibinfo {author} {\bibfnamefont
  {J.~H.}\ \bibnamefont {Eggert}},\ }\href@noop {} {\bibfield  {journal}
  {\bibinfo  {journal} {Nature}\ }\textbf {\bibinfo {volume} {569}},\ \bibinfo
  {pages} {251} (\bibinfo {year} {2019})}\BibitemShut {NoStop}%
\bibitem [{\citenamefont {Chen}\ \emph {et~al.}(2018)\citenamefont {Chen},
  \citenamefont {Zheng}, \citenamefont {Santra}, \citenamefont {Ko},
  \citenamefont {DiStasio~Jr}, \citenamefont {Klein}, \citenamefont {Car},\
  and\ \citenamefont {Wu}}]{chen2018hydroxide}%
  \BibitemOpen
  \bibfield  {author} {\bibinfo {author} {\bibfnamefont {M.}~\bibnamefont
  {Chen}}, \bibinfo {author} {\bibfnamefont {L.}~\bibnamefont {Zheng}},
  \bibinfo {author} {\bibfnamefont {B.}~\bibnamefont {Santra}}, \bibinfo
  {author} {\bibfnamefont {H.-Y.}\ \bibnamefont {Ko}}, \bibinfo {author}
  {\bibfnamefont {R.~A.}\ \bibnamefont {DiStasio~Jr}}, \bibinfo {author}
  {\bibfnamefont {M.~L.}\ \bibnamefont {Klein}}, \bibinfo {author}
  {\bibfnamefont {R.}~\bibnamefont {Car}}, \ and\ \bibinfo {author}
  {\bibfnamefont {X.}~\bibnamefont {Wu}},\ }\href@noop {} {\bibfield  {journal}
  {\bibinfo  {journal} {Nature chemistry}\ }\textbf {\bibinfo {volume} {10}},\
  \bibinfo {pages} {413} (\bibinfo {year} {2018})}\BibitemShut {NoStop}%
\bibitem [{\citenamefont {Kingma}\ and\ \citenamefont
  {Ba}(2015)}]{Kingma2015adam}%
  \BibitemOpen
  \bibfield  {author} {\bibinfo {author} {\bibfnamefont {D.}~\bibnamefont
  {Kingma}}\ and\ \bibinfo {author} {\bibfnamefont {J.}~\bibnamefont {Ba}},\
  }in\ \href@noop {} {\emph {\bibinfo {booktitle} {Proceedings of the
  International Conference on Learning Representations (ICLR)}}}\ (\bibinfo
  {year} {2015})\BibitemShut {NoStop}%
\bibitem [{\citenamefont {Sommers}\ \emph {et~al.}(2020)\citenamefont
  {Sommers}, \citenamefont {Calegari~Andrade}, \citenamefont {Zhang},
  \citenamefont {Wang},\ and\ \citenamefont {Car}}]{grace2020raman}%
  \BibitemOpen
  \bibfield  {author} {\bibinfo {author} {\bibfnamefont {G.~M.}\ \bibnamefont
  {Sommers}}, \bibinfo {author} {\bibfnamefont {M.~F.}\ \bibnamefont
  {Calegari~Andrade}}, \bibinfo {author} {\bibfnamefont {L.}~\bibnamefont
  {Zhang}}, \bibinfo {author} {\bibfnamefont {H.}~\bibnamefont {Wang}}, \ and\
  \bibinfo {author} {\bibfnamefont {R.}~\bibnamefont {Car}},\ }\href {\doibase
  10.1039/D0CP01893G} {\bibfield  {journal} {\bibinfo  {journal} {Phys. Chem.
  Chem. Phys.}\ }\textbf {\bibinfo {volume} {22}},\ \bibinfo {pages} {10592}
  (\bibinfo {year} {2020})}\BibitemShut {NoStop}%
\bibitem [{\citenamefont {Zhang}\ \emph {et~al.}(2019)\citenamefont {Zhang},
  \citenamefont {Lin}, \citenamefont {Wang}, \citenamefont {Car},\ and\
  \citenamefont {E}}]{zhang2019active}%
  \BibitemOpen
  \bibfield  {author} {\bibinfo {author} {\bibfnamefont {L.}~\bibnamefont
  {Zhang}}, \bibinfo {author} {\bibfnamefont {D.-Y.}\ \bibnamefont {Lin}},
  \bibinfo {author} {\bibfnamefont {H.}~\bibnamefont {Wang}}, \bibinfo {author}
  {\bibfnamefont {R.}~\bibnamefont {Car}}, \ and\ \bibinfo {author}
  {\bibfnamefont {W.}~\bibnamefont {E}},\ }\href@noop {} {\bibfield  {journal}
  {\bibinfo  {journal} {Physical Review Materials}\ }\textbf {\bibinfo {volume}
  {3}},\ \bibinfo {pages} {023804} (\bibinfo {year} {2019})}\BibitemShut
  {NoStop}%
\bibitem [{\citenamefont {Podryabinkin}\ and\ \citenamefont
  {Shapeev}(2017)}]{podryabinkin2017active}%
  \BibitemOpen
  \bibfield  {author} {\bibinfo {author} {\bibfnamefont {E.~V.}\ \bibnamefont
  {Podryabinkin}}\ and\ \bibinfo {author} {\bibfnamefont {A.~V.}\ \bibnamefont
  {Shapeev}},\ }\href@noop {} {\bibfield  {journal} {\bibinfo  {journal}
  {Computational Materials Science}\ }\textbf {\bibinfo {volume} {140}},\
  \bibinfo {pages} {171} (\bibinfo {year} {2017})}\BibitemShut {NoStop}%
\bibitem [{\citenamefont {Zhang}\ \emph
  {et~al.}(2018{\natexlab{c}})\citenamefont {Zhang}, \citenamefont {Wang},\
  and\ \citenamefont {E}}]{zhang2018reinforced}%
  \BibitemOpen
  \bibfield  {author} {\bibinfo {author} {\bibfnamefont {L.}~\bibnamefont
  {Zhang}}, \bibinfo {author} {\bibfnamefont {H.}~\bibnamefont {Wang}}, \ and\
  \bibinfo {author} {\bibfnamefont {W.}~\bibnamefont {E}},\ }\href@noop {}
  {\bibfield  {journal} {\bibinfo  {journal} {The Journal of Chemical Physics}\
  }\textbf {\bibinfo {volume} {148}},\ \bibinfo {pages} {124113} (\bibinfo
  {year} {2018}{\natexlab{c}})}\BibitemShut {NoStop}%
\bibitem [{\citenamefont {Ko}\ \emph {et~al.}(2019)\citenamefont {Ko},
  \citenamefont {Zhang}, \citenamefont {Santra}, \citenamefont {Wang},
  \citenamefont {E}, \citenamefont {DiStasio~Jr},\ and\ \citenamefont
  {Car}}]{ko2019isotope}%
  \BibitemOpen
  \bibfield  {author} {\bibinfo {author} {\bibfnamefont {H.-Y.}\ \bibnamefont
  {Ko}}, \bibinfo {author} {\bibfnamefont {L.}~\bibnamefont {Zhang}}, \bibinfo
  {author} {\bibfnamefont {B.}~\bibnamefont {Santra}}, \bibinfo {author}
  {\bibfnamefont {H.}~\bibnamefont {Wang}}, \bibinfo {author} {\bibfnamefont
  {W.}~\bibnamefont {E}}, \bibinfo {author} {\bibfnamefont {R.~A.}\
  \bibnamefont {DiStasio~Jr}}, \ and\ \bibinfo {author} {\bibfnamefont
  {R.}~\bibnamefont {Car}},\ }\href@noop {} {\bibfield  {journal} {\bibinfo
  {journal} {Molecular Physics}\ ,\ \bibinfo {pages} {1}} (\bibinfo {year}
  {2019})}\BibitemShut {NoStop}%
\bibitem [{\citenamefont {Wang}\ \emph {et~al.}(2018)\citenamefont {Wang},
  \citenamefont {Zhang}, \citenamefont {Han},\ and\ \citenamefont
  {E}}]{wang2018kit}%
  \BibitemOpen
  \bibfield  {author} {\bibinfo {author} {\bibfnamefont {H.}~\bibnamefont
  {Wang}}, \bibinfo {author} {\bibfnamefont {L.}~\bibnamefont {Zhang}},
  \bibinfo {author} {\bibfnamefont {J.}~\bibnamefont {Han}}, \ and\ \bibinfo
  {author} {\bibfnamefont {W.}~\bibnamefont {E}},\ }\href@noop {} {\bibfield
  {journal} {\bibinfo  {journal} {Computer Physics Communications}\ }\textbf
  {\bibinfo {volume} {228}},\ \bibinfo {pages} {178 } (\bibinfo {year}
  {2018})}\BibitemShut {NoStop}%
\bibitem [{\citenamefont {Zhang}\ \emph {et~al.}(2020)\citenamefont {Zhang},
  \citenamefont {Wang}, \citenamefont {Chen}, \citenamefont {Zeng},
  \citenamefont {Zhang}, \citenamefont {Wang},\ and\ \citenamefont
  {E}}]{zhang2019dpgen}%
  \BibitemOpen
  \bibfield  {author} {\bibinfo {author} {\bibfnamefont {Y.}~\bibnamefont
  {Zhang}}, \bibinfo {author} {\bibfnamefont {H.}~\bibnamefont {Wang}},
  \bibinfo {author} {\bibfnamefont {W.}~\bibnamefont {Chen}}, \bibinfo {author}
  {\bibfnamefont {J.}~\bibnamefont {Zeng}}, \bibinfo {author} {\bibfnamefont
  {L.}~\bibnamefont {Zhang}}, \bibinfo {author} {\bibfnamefont
  {H.}~\bibnamefont {Wang}}, \ and\ \bibinfo {author} {\bibfnamefont
  {W.}~\bibnamefont {E}},\ }\href@noop {} {\bibfield  {journal} {\bibinfo
  {journal} {Computer Physics Communications}\ ,\ \bibinfo {pages} {107206}}
  (\bibinfo {year} {2020})}\BibitemShut {NoStop}%
\bibitem [{\citenamefont {Bertie}\ and\ \citenamefont
  {Lan}(1996)}]{bertie1996infrared}%
  \BibitemOpen
  \bibfield  {author} {\bibinfo {author} {\bibfnamefont {J.~E.}\ \bibnamefont
  {Bertie}}\ and\ \bibinfo {author} {\bibfnamefont {Z.}~\bibnamefont {Lan}},\
  }\href@noop {} {\bibfield  {journal} {\bibinfo  {journal} {Applied
  Spectroscopy}\ }\textbf {\bibinfo {volume} {50}},\ \bibinfo {pages} {1047}
  (\bibinfo {year} {1996})}\BibitemShut {NoStop}%
\bibitem [{\citenamefont {Bertie}\ \emph {et~al.}(1989)\citenamefont {Bertie},
  \citenamefont {Ahmed},\ and\ \citenamefont {Eysel}}]{bertie1989infrared}%
  \BibitemOpen
  \bibfield  {author} {\bibinfo {author} {\bibfnamefont {J.~E.}\ \bibnamefont
  {Bertie}}, \bibinfo {author} {\bibfnamefont {M.~K.}\ \bibnamefont {Ahmed}}, \
  and\ \bibinfo {author} {\bibfnamefont {H.~H.}\ \bibnamefont {Eysel}},\
  }\href@noop {} {\bibfield  {journal} {\bibinfo  {journal} {The Journal of
  Physical Chemistry}\ }\textbf {\bibinfo {volume} {93}},\ \bibinfo {pages}
  {2210} (\bibinfo {year} {1989})}\BibitemShut {NoStop}%
\bibitem [{\citenamefont {Adamo}\ and\ \citenamefont
  {Barone}(1999)}]{Carlo1999PBE0}%
  \BibitemOpen
  \bibfield  {author} {\bibinfo {author} {\bibfnamefont {C.}~\bibnamefont
  {Adamo}}\ and\ \bibinfo {author} {\bibfnamefont {V.}~\bibnamefont {Barone}},\
  }\href@noop {} {\bibfield  {journal} {\bibinfo  {journal} {The Journal of
  Chemical Physics}\ }\textbf {\bibinfo {volume} {110}},\ \bibinfo {pages}
  {6158} (\bibinfo {year} {1999})}\BibitemShut {NoStop}%
\bibitem [{\citenamefont {Tkatchenko}\ and\ \citenamefont
  {Scheffler}(2009)}]{TS2009TS}%
  \BibitemOpen
  \bibfield  {author} {\bibinfo {author} {\bibfnamefont {A.}~\bibnamefont
  {Tkatchenko}}\ and\ \bibinfo {author} {\bibfnamefont {M.}~\bibnamefont
  {Scheffler}},\ }\href@noop {} {\bibfield  {journal} {\bibinfo  {journal}
  {Physical Review Letters}\ }\textbf {\bibinfo {volume} {102}},\ \bibinfo
  {pages} {073005} (\bibinfo {year} {2009})}\BibitemShut {NoStop}%
\bibitem [{\citenamefont {Babin}\ \emph {et~al.}(2013)\citenamefont {Babin},
  \citenamefont {Leforestier},\ and\ \citenamefont
  {Paesani}}]{babin2013development}%
  \BibitemOpen
  \bibfield  {author} {\bibinfo {author} {\bibfnamefont {V.}~\bibnamefont
  {Babin}}, \bibinfo {author} {\bibfnamefont {C.}~\bibnamefont {Leforestier}},
  \ and\ \bibinfo {author} {\bibfnamefont {F.}~\bibnamefont {Paesani}},\
  }\href@noop {} {\bibfield  {journal} {\bibinfo  {journal} {Journal of
  Chemical Theory and Computation}\ }\textbf {\bibinfo {volume} {9}},\ \bibinfo
  {pages} {5395} (\bibinfo {year} {2013})}\BibitemShut {NoStop}%
\bibitem [{\citenamefont {Shank}\ \emph {et~al.}(2009)\citenamefont {Shank},
  \citenamefont {Wang}, \citenamefont {Kaledin}, \citenamefont {Braams},\ and\
  \citenamefont {Bowman}}]{shank2009accurate}%
  \BibitemOpen
  \bibfield  {author} {\bibinfo {author} {\bibfnamefont {A.}~\bibnamefont
  {Shank}}, \bibinfo {author} {\bibfnamefont {Y.}~\bibnamefont {Wang}},
  \bibinfo {author} {\bibfnamefont {A.}~\bibnamefont {Kaledin}}, \bibinfo
  {author} {\bibfnamefont {B.~J.}\ \bibnamefont {Braams}}, \ and\ \bibinfo
  {author} {\bibfnamefont {J.~M.}\ \bibnamefont {Bowman}},\ }\href@noop {}
  {\bibfield  {journal} {\bibinfo  {journal} {The Journal of Chemical Physics}\
  }\textbf {\bibinfo {volume} {130}},\ \bibinfo {pages} {144314} (\bibinfo
  {year} {2009})}\BibitemShut {NoStop}%
\bibitem [{\citenamefont {Nos{\'e}}(1984)}]{nose1984unified}%
  \BibitemOpen
  \bibfield  {author} {\bibinfo {author} {\bibfnamefont {S.}~\bibnamefont
  {Nos{\'e}}},\ }\href@noop {} {\bibfield  {journal} {\bibinfo  {journal} {The
  Journal of Chemical Physics}\ }\textbf {\bibinfo {volume} {81}},\ \bibinfo
  {pages} {511} (\bibinfo {year} {1984})}\BibitemShut {NoStop}%
\bibitem [{\citenamefont {Hoover}(1985)}]{hoover1985canonical}%
  \BibitemOpen
  \bibfield  {author} {\bibinfo {author} {\bibfnamefont {W.~G.}\ \bibnamefont
  {Hoover}},\ }\href@noop {} {\bibfield  {journal} {\bibinfo  {journal}
  {Physical review A}\ }\textbf {\bibinfo {volume} {31}},\ \bibinfo {pages}
  {1695} (\bibinfo {year} {1985})}\BibitemShut {NoStop}%
\bibitem [{\citenamefont {Bernasconi}\ \emph {et~al.}(1998)\citenamefont
  {Bernasconi}, \citenamefont {Silvestrelli},\ and\ \citenamefont
  {Parrinello}}]{bernasconi1998ab}%
  \BibitemOpen
  \bibfield  {author} {\bibinfo {author} {\bibfnamefont {M.}~\bibnamefont
  {Bernasconi}}, \bibinfo {author} {\bibfnamefont {P.}~\bibnamefont
  {Silvestrelli}}, \ and\ \bibinfo {author} {\bibfnamefont {M.}~\bibnamefont
  {Parrinello}},\ }\href@noop {} {\bibfield  {journal} {\bibinfo  {journal}
  {Physical Review Letters}\ }\textbf {\bibinfo {volume} {81}},\ \bibinfo
  {pages} {1235} (\bibinfo {year} {1998})}\BibitemShut {NoStop}%
\bibitem [{\citenamefont {Goncharov}\ \emph {et~al.}(1996)\citenamefont
  {Goncharov}, \citenamefont {Struzhkin}, \citenamefont {Somayazulu},
  \citenamefont {Hemley},\ and\ \citenamefont
  {Mao}}]{goncharov1996compression}%
  \BibitemOpen
  \bibfield  {author} {\bibinfo {author} {\bibfnamefont {A.}~\bibnamefont
  {Goncharov}}, \bibinfo {author} {\bibfnamefont {V.}~\bibnamefont
  {Struzhkin}}, \bibinfo {author} {\bibfnamefont {M.}~\bibnamefont
  {Somayazulu}}, \bibinfo {author} {\bibfnamefont {R.}~\bibnamefont {Hemley}},
  \ and\ \bibinfo {author} {\bibfnamefont {H.}~\bibnamefont {Mao}},\
  }\href@noop {} {\bibfield  {journal} {\bibinfo  {journal} {Science}\ }\textbf
  {\bibinfo {volume} {273}},\ \bibinfo {pages} {218} (\bibinfo {year}
  {1996})}\BibitemShut {NoStop}%
\bibitem [{\citenamefont {Marsalek}\ and\ \citenamefont
  {Markland}(2017)}]{marsalek2017quantum}%
  \BibitemOpen
  \bibfield  {author} {\bibinfo {author} {\bibfnamefont {O.}~\bibnamefont
  {Marsalek}}\ and\ \bibinfo {author} {\bibfnamefont {T.~E.}\ \bibnamefont
  {Markland}},\ }\href@noop {} {\bibfield  {journal} {\bibinfo  {journal} {The
  Journal of Physical Chemistry Letters}\ }\textbf {\bibinfo {volume} {8}},\
  \bibinfo {pages} {1545} (\bibinfo {year} {2017})}\BibitemShut {NoStop}%
\bibitem [{\citenamefont {Medders}\ and\ \citenamefont
  {Paesani}(2015)}]{medders2015infrared}%
  \BibitemOpen
  \bibfield  {author} {\bibinfo {author} {\bibfnamefont {G.~R.}\ \bibnamefont
  {Medders}}\ and\ \bibinfo {author} {\bibfnamefont {F.}~\bibnamefont
  {Paesani}},\ }\href@noop {} {\bibfield  {journal} {\bibinfo  {journal}
  {Journal of Chemical Theory and Computation}\ }\textbf {\bibinfo {volume}
  {11}},\ \bibinfo {pages} {1145} (\bibinfo {year} {2015})}\BibitemShut
  {NoStop}%
\bibitem [{\citenamefont {Mukamel}(1998)}]{Mukamel1998Principles}%
  \BibitemOpen
  \bibfield  {author} {\bibinfo {author} {\bibfnamefont {S.}~\bibnamefont
  {Mukamel}},\ }\href@noop {} {\emph {\bibinfo {title} {Principles of nonlinear
  optical spectroscopy}}},\ Vol.~\bibinfo {volume} {41}\ (\bibinfo {year}
  {1998})\ pp.\ \bibinfo {pages} {591--592}\BibitemShut {NoStop}%
\bibitem [{\citenamefont {Nagata}\ and\ \citenamefont
  {Mukamel}(2010)}]{Nagata2010Vibrational}%
  \BibitemOpen
  \bibfield  {author} {\bibinfo {author} {\bibfnamefont {Y.}~\bibnamefont
  {Nagata}}\ and\ \bibinfo {author} {\bibfnamefont {S.}~\bibnamefont
  {Mukamel}},\ }\href@noop {} {\bibfield  {journal} {\bibinfo  {journal}
  {Journal of the American Chemical Society}\ }\textbf {\bibinfo {volume}
  {132}},\ \bibinfo {pages} {6434} (\bibinfo {year} {2010})}\BibitemShut
  {NoStop}%
\bibitem [{\citenamefont {Resta}\ and\ \citenamefont
  {Vanderbilt}(2007)}]{resta2007theory}%
  \BibitemOpen
  \bibfield  {author} {\bibinfo {author} {\bibfnamefont {R.}~\bibnamefont
  {Resta}}\ and\ \bibinfo {author} {\bibfnamefont {D.}~\bibnamefont
  {Vanderbilt}},\ }in\ \href@noop {} {\emph {\bibinfo {booktitle} {Physics of
  Ferroelectrics}}}\ (\bibinfo  {publisher} {Springer},\ \bibinfo {year}
  {2007})\ pp.\ \bibinfo {pages} {31--68}\BibitemShut {NoStop}%
\bibitem [{\citenamefont {Vanderbilt}\ and\ \citenamefont
  {Zhong}(1998)}]{david1998first}%
  \BibitemOpen
  \bibfield  {author} {\bibinfo {author} {\bibfnamefont {D.}~\bibnamefont
  {Vanderbilt}}\ and\ \bibinfo {author} {\bibfnamefont {W.}~\bibnamefont
  {Zhong}},\ }\href@noop {} {\bibfield  {journal} {\bibinfo  {journal}
  {Ferroelectrics}\ }\textbf {\bibinfo {volume} {206}},\ \bibinfo {pages} {181}
  (\bibinfo {year} {1998})}\BibitemShut {NoStop}%
\bibitem [{\citenamefont {Zaheer}\ \emph {et~al.}(2017)\citenamefont {Zaheer},
  \citenamefont {Kottur}, \citenamefont {Ravanbakhsh}, \citenamefont {Poczos},
  \citenamefont {Salakhutdinov},\ and\ \citenamefont
  {Smola}}]{Zaheer2017deepsets}%
  \BibitemOpen
  \bibfield  {author} {\bibinfo {author} {\bibfnamefont {M.}~\bibnamefont
  {Zaheer}}, \bibinfo {author} {\bibfnamefont {S.}~\bibnamefont {Kottur}},
  \bibinfo {author} {\bibfnamefont {S.}~\bibnamefont {Ravanbakhsh}}, \bibinfo
  {author} {\bibfnamefont {B.}~\bibnamefont {Poczos}}, \bibinfo {author}
  {\bibfnamefont {R.~R.}\ \bibnamefont {Salakhutdinov}}, \ and\ \bibinfo
  {author} {\bibfnamefont {A.~J.}\ \bibnamefont {Smola}},\ }in\ \href
  {http://papers.nips.cc/paper/6931-deep-sets.pdf} {\emph {\bibinfo {booktitle}
  {Advances in Neural Information Processing Systems 30}}},\ \bibinfo {editor}
  {edited by\ \bibinfo {editor} {\bibfnamefont {I.}~\bibnamefont {Guyon}},
  \bibinfo {editor} {\bibfnamefont {U.~V.}\ \bibnamefont {Luxburg}}, \bibinfo
  {editor} {\bibfnamefont {S.}~\bibnamefont {Bengio}}, \bibinfo {editor}
  {\bibfnamefont {H.}~\bibnamefont {Wallach}}, \bibinfo {editor} {\bibfnamefont
  {R.}~\bibnamefont {Fergus}}, \bibinfo {editor} {\bibfnamefont
  {S.}~\bibnamefont {Vishwanathan}}, \ and\ \bibinfo {editor} {\bibfnamefont
  {R.}~\bibnamefont {Garnett}}}\ (\bibinfo  {publisher} {Curran Associates,
  Inc.},\ \bibinfo {year} {2017})\ pp.\ \bibinfo {pages}
  {3391--3401}\BibitemShut {NoStop}%
\bibitem [{\citenamefont {Battaglia}\ \emph {et~al.}(2018)\citenamefont
  {Battaglia}, \citenamefont {Hamrick}, \citenamefont {Bapst}, \citenamefont
  {Sanchez-Gonzalez}, \citenamefont {Zambaldi}, \citenamefont {Malinowski},
  \citenamefont {Tacchetti}, \citenamefont {Raposo}, \citenamefont {Santoro},
  \citenamefont {Faulkner} \emph {et~al.}}]{battaglia2018relational}%
  \BibitemOpen
  \bibfield  {author} {\bibinfo {author} {\bibfnamefont {P.~W.}\ \bibnamefont
  {Battaglia}}, \bibinfo {author} {\bibfnamefont {J.~B.}\ \bibnamefont
  {Hamrick}}, \bibinfo {author} {\bibfnamefont {V.}~\bibnamefont {Bapst}},
  \bibinfo {author} {\bibfnamefont {A.}~\bibnamefont {Sanchez-Gonzalez}},
  \bibinfo {author} {\bibfnamefont {V.}~\bibnamefont {Zambaldi}}, \bibinfo
  {author} {\bibfnamefont {M.}~\bibnamefont {Malinowski}}, \bibinfo {author}
  {\bibfnamefont {A.}~\bibnamefont {Tacchetti}}, \bibinfo {author}
  {\bibfnamefont {D.}~\bibnamefont {Raposo}}, \bibinfo {author} {\bibfnamefont
  {A.}~\bibnamefont {Santoro}}, \bibinfo {author} {\bibfnamefont
  {R.}~\bibnamefont {Faulkner}},  \emph {et~al.},\ }\href@noop {} {\bibfield
  {journal} {\bibinfo  {journal} {arXiv preprint arXiv:1806.01261}\ } (\bibinfo
  {year} {2018})}\BibitemShut {NoStop}%
\bibitem [{\citenamefont {Esteves}\ \emph {et~al.}(2018)\citenamefont
  {Esteves}, \citenamefont {Allen-Blanchette}, \citenamefont {Makadia},\ and\
  \citenamefont {Daniilidis}}]{esteves2018learning}%
  \BibitemOpen
  \bibfield  {author} {\bibinfo {author} {\bibfnamefont {C.}~\bibnamefont
  {Esteves}}, \bibinfo {author} {\bibfnamefont {C.}~\bibnamefont
  {Allen-Blanchette}}, \bibinfo {author} {\bibfnamefont {A.}~\bibnamefont
  {Makadia}}, \ and\ \bibinfo {author} {\bibfnamefont {K.}~\bibnamefont
  {Daniilidis}},\ }in\ \href@noop {} {\emph {\bibinfo {booktitle} {Proceedings
  of the European Conference on Computer Vision (ECCV)}}}\ (\bibinfo {year}
  {2018})\ pp.\ \bibinfo {pages} {52--68}\BibitemShut {NoStop}%
\bibitem [{\citenamefont {Han}\ \emph {et~al.}(2019)\citenamefont {Han},
  \citenamefont {Li}, \citenamefont {Lin}, \citenamefont {Lu}, \citenamefont
  {Zhang},\ and\ \citenamefont {Zhang}}]{han2019universal}%
  \BibitemOpen
  \bibfield  {author} {\bibinfo {author} {\bibfnamefont {J.}~\bibnamefont
  {Han}}, \bibinfo {author} {\bibfnamefont {Y.}~\bibnamefont {Li}}, \bibinfo
  {author} {\bibfnamefont {L.}~\bibnamefont {Lin}}, \bibinfo {author}
  {\bibfnamefont {J.}~\bibnamefont {Lu}}, \bibinfo {author} {\bibfnamefont
  {J.}~\bibnamefont {Zhang}}, \ and\ \bibinfo {author} {\bibfnamefont
  {L.}~\bibnamefont {Zhang}},\ }\href@noop {} {\bibfield  {journal} {\bibinfo
  {journal} {arXiv preprint arXiv:1912.01765}\ } (\bibinfo {year}
  {2019})}\BibitemShut {NoStop}%
\end{thebibliography}
\end{document}